\documentclass[twocolumn,tighten,times,numberedappendix,twocolappendix]{aastex6}
\usepackage{color}
\usepackage{graphicx}
\usepackage{mathrsfs}
\usepackage{ulem}
\usepackage{empheq}
\usepackage{multirow}
\usepackage{array}

\DeclareMathAlphabet{\mathbi}{OT1}{ptm}{bx}{it}
\SetMathAlphabet\mathbi{bold}{OT1}{ptm}{bx}{it}

\def\bhm{M_{\bullet}}

\def\ergs{\rm erg~s^{-1}}

\def\mathdotM{\dot{\mathscr{M}}}

\def\sunm{M_\odot}

\def\mathdotM{\dot{\mathscr{M}}}

\begin{document}

\title{Active Galactic Nuclei with Ultra-fast Outflows Monitoring Project: The Broad-line Region of Mrk~79 as a Disk Wind} 

\author{Kai-Xing Lu\altaffilmark{1,2,11}, 
Jin-Ming Bai\altaffilmark{1,2,11}, 
Zhi-Xiang Zhang\altaffilmark{3}, 
Pu Du\altaffilmark{4}, 
Chen Hu\altaffilmark{4}, 
Minjin Kim\altaffilmark{5,6}, 
Jian-Min Wang\altaffilmark{4,11}, 
Luis C. Ho\altaffilmark{7,8}, 
Yan-Rong Li\altaffilmark{4}, 
Wei-Hao Bian\altaffilmark{9}, 
Ye-Fei Yuan\altaffilmark{10}, 
Ming Xiao\altaffilmark{4}, 
Hai-Cheng Feng\altaffilmark{1,2},   
Jian-Guo Wang\altaffilmark{1,2},    
Liang Xu\altaffilmark{1,2},               
Xu Ding\altaffilmark{1,2},                 
Xiao-Guang Yu\altaffilmark{1,2},     
Yu-Xin Xin\altaffilmark{1,2},            
Kai Ye\altaffilmark{1,2},                   
Chuan-Jun Wang\altaffilmark{1,2},  
Bao-Li Lun\altaffilmark{1,2},           
Ju-Jia Zhang\altaffilmark{1,2},       
Xi-Liang Zhang\altaffilmark{1,2},    
Kai-Fan Ji\altaffilmark{1}, 
Yu-Feng Fan\altaffilmark{1,2},        
Liang Chang\altaffilmark{1,2}         
}

\altaffiltext{1}{Yunnan Observatories, Chinese Academy of Sciences, Kunming 650011, People's Republic of China}
\altaffiltext{2}{Key Laboratory for the Structure and Evolution of Celestial Objects, Chinese Academy of Sciences, Kunming 650011, People's Republic of China}
\altaffiltext{3}{Department of Astronomy, Xiamen University, Xiamen, Fujian 361005, People's Republic of China}
\altaffiltext{4}{Key Laboratory for Particle Astrophysics, Institute of High Energy Physics, Chinese Academy of Sciences, 19B Yuquan Road, Beijing 100049, People's Republic of China}
\altaffiltext{5}{Department of Astronomy and Atmospheric Sciences, Kyungpook National University, Daegu 702-701, Republic of Korea}
\altaffiltext{6}{Korea Astronomy and Space Science Institute, Daejeon 305-348, Republic of Korea}
\altaffiltext{7}{Kavli Institute for Astronomy and Astrophysics, Peking University, Beijing 100871, People's Republic of China}
\altaffiltext{8}{Department of Astronomy, School of Physics, Peking University, Beijing 100871, People's Republic of China}
\altaffiltext{9}{Physics Department, Nanjing Normal University, Nanjing 210097, People's Republic of China}
\altaffiltext{10}{Department of Astronomy, University of Science and Technology of China, Hefei 230026, People's Republic of China}
\altaffiltext{11}{Corresponding authors: lukx@ynao.ac.cn, baijinming@ynao.ac.cn, wangjm@ihep.ac.cn}

\begin{abstract} 
We developed a spectroscopic monitoring project to investigate the kinematics of the broad-line region (BLR) 
in active galactic nuclei (AGN) with ultra-fast outflows (UFOs). 
Mrk~79 is a radio-quiet AGN with UFOs and warm absorbers, had been monitored by three reverberation mapping (RM) campaigns, 
but its BLR kinematics is not understood yet. 
In this paper, we report the results from a new RM-campaign of Mrk~79, 
which was undertaken by Lijiang 2.4-m telescope. Mrk~79 is seeming to come out the faint state, 
the mean flux approximates a magnitude fainter than historical record. 
We successfully measured the lags of the broad emission lines including H$\beta~\lambda4861$, 
H$\gamma~\lambda4340$, He~{\sc ii}~$\lambda4686$ and He~{\sc i}~$\lambda5876$ with respect to the varying AGN continuum. 
Based on the broad H$\beta~\lambda4861$ line, we measured black hole (BH) mass 
of $M_{\bullet}=5.13^{+1.57}_{-1.55}\times10^{7}M_{\odot}$, 
estimated accretion rates of ${\dot{M}_{\bullet}}=(0.05\pm0.02)~L_{\rm Edd}~c^{-2}$, indicating that Mrk~79 is a sub-Eddington accretor.  
We found that Mrk~79 deviates from the canonical Radius$-$Luminosity relationship. 
The marginal blueshift of the broad He~{\sc ii}~$\lambda4686$ line detected from rms spectrum indicates outflow of high-ionization gas. 
The velocity-resolved lag profiles of the broad H$\gamma~\lambda4340$, H$\beta~\lambda4861$, and He~{\sc i}~$\lambda5876$ lines 
show similar signatures that the largest lag occurs in the red wing of the lines then the lag decreases to both sides. 
These signatures should suggest that the BLR of Keplerian motion probably exists the outflow gas motion. 
All findings including UFOs, warm absorbers, and the kinematics of high- and low-ionization BLR, 
may provide an indirect evidence that the BLR of Mrk~79 probably originates from disk wind. 
\end{abstract}


\section{Introduction}
\label{sec_intro}
In the past thirty  years, reverberation mapping (RM; \citealt{Bahcall1972,Blandford1982}) 
has been extensively adopted to investigate the kinematics of the broad-line region (BLR) 
and measure the mass of accreting supermassive black hole (BH) in active galactic nuclei (AGN). 
BH masses of $\sim$100 AGNs have been measured by different RM campaigns 
(e.g., \citealt{Peterson1993b,Peterson1998,Kaspi2000,Peterson2004,
Denney2009b,Denney2010,Bentz2007,Bentz2009b,Bentz2010,Bentz2013,
Barth2011a,Barth2011b,Peterson2014,Du2015,Grier2017,Du2018,DeRosa2018}). 
Canonical Radius$-$Luminosity relationship is constructed from these RM campaigns 
\citep{Peterson1993b,Wandel1999,Kaspi2000,Bentz2013}, 
which provides an indirect way to estimate BH mass from single spectrum. 
However, as the number of sample increases, this relationship becomes more scatter ($\sim$0.3~dex; \citealt{Bentz2013,Grier2017,Du2018}). 
It is possible to find the physics of this scatter studying the long-term variation of the BLR 
by repeated RM-campaign \citep{Peterson2002,Lu2016}. 
The kinematic structures of the BLR for $\sim$20 AGNs have been probed using velocity-resolved RM 
(e.g., \citealt{Denney2009a,Denney2010,Bentz2009b,Bentz2010,Barth2011a,
Barth2011b,Grier2013,Du2016,Lu2016,Pei2017,DeRosa2018,Zhang2019}). 
They usually include virialized disk, inflow, and outflow \citep{Bentz2009b,Grier2013}, 
but physical origins of inflow and outflow remain unclear. 

Mrk~79 is a nearby and radio-quiet (RQ) AGN ($z=0.022189$ from NED), 
which has been monitored by three RM-campaign \citep{Peterson1998}. 
These campaigns successfully detected time delays of the broad H$\beta$ line with respect to the continuum variation,  
but the BLR's kinematics of Mrk~79 is not understood yet. 
Therefore, Mrk~79 deserves to be monitored again 
to investigate the kinematics of the BLR, and to construct the Radius$-$Luminosity relationship of Mrk~79 
(like Radius$-$Luminosity relationship of NGC~5548, see \citealt{Peterson2002,Lu2016,Pei2017,Kriss2019ApJ}). 

Secondly, based on  X-ray spectrum between 7 and 10 keV, 
ultra-fast outflows (UFOs, i.e., highly ionized absorbers) have been detected in Mrk~79 
through Fe~{\sc xxv} and Fe~{\sc xxvi} K-shell absorption lines with blueshifted velocity $V_{\rm UFOs,out}=(0.092\pm0.004)c$ 
($c$ is the speed of light, see \citealt{Tombesi2010,Tombesi2011}). 
Warm absorbers also detected in soft X-ray spectrum , 
but current energy resolution of X-ray spectrum is hard to constrain its nature \citep{Gallo2011,Tombesi2013}. 
UFOs and warm absorbers are probably associated with accretion disk wind in Mrk~79 because it is a RQ AGN (no jets; \citealt{Tombesi2011}). 
What is the connection between the BLR and accretion disk wind? 

On the one hand, 
multiphase disk wind including hard X-ray absorbers (UFOs), soft X-ray absorbers (warm absorbers), 
and UV absorbers were confirmed in term of the blueshift of X-ray and ultraviolet (UV) broad absorption lines 
(e.g., \citealt{Murray1997,Leighly2004,OBrien2005,Tombesi2010,Tombesi2013,Hamann2018,Longinotti2019,Giustini2019}), 
in which these absorbers jointly or partially exist in AGNs \citep{OBrien2005,Tombesi2013}. 
Therefore a disk wind model characterised by multiphase stratified structure was developed to explain 
these phenomenons (e.g., see Figure 5 of \citealt{Tombesi2013}, Figure 6 of \citealt{Mas-Ribas2019}). 
In addition, outflows of the BLR are also found from UV and optical spectra. 
For example, (1) \cite{Richards2011} found that the blueshift of broad C~{\sc iv} emission line are nearly ubiquitous, 
with a mean velocity of $\sim810~\rm km~s^{-1}$ for RQ AGNs; 
(2) in RM study of spectroscopic monitoring, \cite{Hu2015} found that a blueshifted 
broad He~{\sc ii}~$\lambda4686$ line is need to reasonably decompose optical spectrum; 
(3) based on the rest frame defined by [O~{\sc iii}]~$\lambda5007$ line, 
\cite{Ge2019} found the blueshift of the broad C~{\sc iv} emission line 
has a medium-strong positive correlation with the optical luminosity and the Eddington ratio. 
\cite{Murray1995} developed a model for line-driven wind from accretion disk, which suggested that 
the absorbing gas cannot lie within the broad-line emitting region but can be cospatial with it or outside of it, 
and predicted that the high-ionization emission lines should be blueshifted relative to the low-ionization emission lines. 
Interestingly, the blueshift of absorbers and emitters jointly exist in a few AGNs (e.g., PDS~ 456). 
\cite{OBrien2005} attributed this phenomenon to a decelerating, cooling outflow, which may be driven 
by radiation and/or magnetic field, and suggested that the X-ray outflow could be the source of some of the BLR gas. 
In this case, 
the geometric and kinematic structures of the BLR could be modified by decelerating, cooling outflow.  

On the other hand, 
The signature of the outflowing BLR was just observed doubtlessly by velocity-resolved RM in Mrk~142 and NGC~3227 
\citep{Du2016,Denney2009a,Denney2010}. 
It's worth noting that these both AGNs have different properties. 
Mrk~142 has very high accretion rates, radiation pressure acting on the ionized gas 
may drive outflow of the BLR \citep{Du2015}. 
However, NGC~3227 is a low-accretion AGN \citep{Denney2010,Du2015}, 
but has detected outflows of hard X-ray and soft X-ray absorbers with velocity of 
$\sim \rm 2060~km~s^{-1}$ and $\sim \rm 420~km~s^{-1}$ (i.e., UFOs and warm absorbers) 
from X-ray spectrum \citep{Markowitz2009}. 
What drives outflow of the BLR for AGNs in the low accretion rate? 

Motivated by above questions, 
we will focus on investigating the BLR kinematics of AGNs with 
UFOs and explore potential connection between the BLR and UFOs (or accretion disk wind). 
Therefore we developed an AGNs with UFOs monitoring project, 
this paper presents the results from the spectroscopic monitoring of Mrk~79. 
The paper is organized as follows. 
In Section~2, we describe the observation and the data reduction in detail. 
Data analysis and results including spectral measurement, time series analysis, 
the construction of velocity-resolved lag profiles 
and estimate black hole mass and so on are present in Section~3. 
Section~4 is discussion and summary is given in Section~5. 
We use a cosmology with $H_0=67{\rm~km~s^{-1}~Mpc^{-1}}$, $\Omega_{\Lambda}=0.68$, 
and $\Omega_{\rm M}=0.32$ \citep{Planck2014}, 

\section{Observation and Data Reduction}
\label{sec_od}
\subsection{Spectroscopic and Photometric Observation}
\label{sec_obs}
The spectroscopic and photometric observation of Mrk~79 were taken using Yunnan Faint Object Spectrograph and Camera (YFOSC) 
mounted on the Lijiang 2.4-m telescope (LJT), 
which is located at Lijiang observatory and is administered by Yunnan Observatories of Chinese Academy of Sciences \citep{Fan2015,Wang2019}. 
YFOSC is equipped with a  back-illuminated 2048$\times$2048 pixel CCD, 
with pixel size 13.5 $\mu$m, pixel scale 0.283$''$ per pixel, and field-of-view $10'\times10'$. 
It is a versatile instrument for low-resolution spectroscopy and photometry (see \citealt{Du2014,Lu2016}). 

The spectral monitoring of Mrk 79 started on 2017 November 1, and terminated on 2018 March 13. 
During the spectroscopic observation, we simultaneously observed a nearby comparison star along 
the slit as a reference standard, which can provide high-precision flux calibration (see \citealt{Hu2015,Lu2016,Du2018}). 
This observation method was described in detail by \citet{Maoz1990} 
and \citet{Kaspi2000}, and was recently adopted by \citep{Du2014,Lu2016,Du2018}. 
In the light of the average seeing $\sim1.3^{\prime\prime}$ of observational station, 
we fixed the projected slit width $2.5''$.  We used Grism 14, 
which provides a resolution 92 \AA\, mm$^{-1}$ (1.8 \AA\, pixel$^{-1}$) and covers the wavelength range 3600$-$7460 \AA\,. 
Standard neon and helium lamps were used for wavelength calibration. 
In total, we obtained 72 spectroscopic observations, spanning a observation period of 132 days. 
The median and mean cadences are 1.0 and 1.8 days, respectively. 
All spectra were observed with a median air mass of 1.15, 
which means that atmospheric differential refraction has a negligible impact on our analysis \citep{Filippenko1982}. 

It should be noted that the standard spectral calibration method assumes that 
the [O~{\sc iii}]~$\lambda5007$ flux is constant and use it as an internal flux calibrator (e.g., \citealt{Fausnaugh2017}). 
To integrally put extended [O~{\sc iii}]~$\lambda5007$ emission region into the slit in spectroscopy, 
many previous RM campaigns adopted a broad slit ($\sim 5^{\prime\prime}$; see Table 12 of \citealt{Bentz2013}) 
at the cost of losing the spectral resolution. In this case, standard spectral calibration method using the 
[O~{\sc iii}]~$\lambda5007$ as calibrator provides precise internal flux calibration of spectra (see \citealt{Fausnaugh2017}). 
Actually, a relatively narrow slit contributes to increasing the spectral resolution, and high spectral resolution is conducive to the 
following velocity-resolved time series analysis (Section~\ref{sec_vm}). 
Therefore, we adopted a relatively narrow slit ($2.5^{\prime\prime}$) in spectroscopy. 
For a small spectrograph slit, 
using the [O~{\sc iii}]~$\lambda5007$ as an internal flux calibrator is not an optimal choice, 
because the [O~{\sc iii}]~$\lambda5007$ emission region of many AGNs 
(e.g., \citealt{Schmitt2003a,Schmitt2003b}, and \citealt{Peterson1995} for NGC~4151) 
along with the host galaxy are (slightly and very) extended sources (see Appendix~\ref{sec_a}), 
varying observing conditions may cause the apparent variation in flux of extended sources (details refer to Appendix~\ref{sec_a}). 
While, in this case, a stable comparison star observed along with the object simultaneously in a narrow slit 
can provide precise flux calibration of spectra (see Section~\ref{sec_lc} and Appendix~\ref{sec_a}, \citealt{Hu2015} as well). 

The photometric images with the field of $10^{\prime\prime}\times10^{\prime\prime}$ 
were obtained using a Johnson $V$ filter. Totally, we obtained 62 photometric observations. 
Typically, two exposures of 90~s were taken for each individual observation. 

\begin{figure}[ht!]
\centering
\includegraphics[angle=0,width=0.5\textwidth]{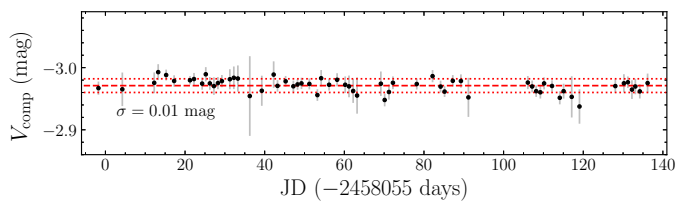}
\caption{\footnotesize
Light curves of the comparison star for the present campaign. 
The scatter of the light curves is 0.01~mag. 
}
\label{fig_complc}
\end{figure}

\subsection{Data reduction}
\label{sec_dr}
The photometric images were reduced following standard {\tt IRAF} procedures using {\tt IRAF(v2.16)} package. 
The magnitude of the object (Mrk~79) and the comparison star were measured through 
a circular aperture with radius of $5.7^{\prime\prime}$, 
and differential magnitudes were obtained relative to 4 selected stars within the field of view. 
Figure~\ref{fig_complc} shows the light curve of comparison star. The accurate of the photometry is 1\%, 
which demonstrates that the comparison star is stable enough to be used for flux calibration of the spectra. 
The photometric data of Mrk~79 will be used to check the spectral calibration in Section~\ref{sec_lc}. 

The two-dimensional spectroscopic data were reduced using the standard {\tt IRAF(v2.16)} package. 
This process included bias subtraction, flat-field correction, wavelength calibration and spectrum extraction. 
All spectra were extracted using a uniform aperture of 20 pixels (5.7$^{\prime\prime}$), 
and background was determined from two adjacent regions ($+7.4^{\prime\prime}\sim+14^{\prime\prime}$ 
and $-7.4^{\prime\prime}\sim-14^{\prime\prime}$) on both sides of the aperture region. 
Actually, a relatively small extraction aperture contributes to reducing the poisson noise of sky background and increasing the signal-to-noise ratio (S/N) of spectrum. 
High S/N ratio is conducive to the following multi-component decomposition of spectrum (Section~\ref{sec_fit}). 
Spectral flux of the target were calibrated by the comparison stars in two steps. 
(1) We produced the fiducial spectrum of the comparison star using data from nights with photometric conditions. 
(2) For each object/comparison star pair, we obtained a wavelength-dependent sensitivity function comparing 
the star's spectrum to the fiducial spectrum. 
Then this sensitivity function was applied to calibrate the observed spectrum of the target 
(also see Appendix~\ref{sec_a}; \citealt{Du2014,Lu2016}). 

\subsection{Data processing} 
\label{sec_dp}
The flux-calibration spectra were corrected for Galactic extinction using the extinction map of \cite{Schlegel1998} at first. 
The variations of seeing and mis-centering usually cause slight wavelength shift 
and broadening of emission lines \citep{Du2018,Lu2016}. 
We corrected wavelength shift using [O~{\sc iii}]~$\lambda5007$ line as wavelength reference, 
and corrected broadening of emission lines convoluting [O~{\sc iii}]~$\lambda5007$ 
line into its maximum width determined from 72~spectra. 
Then spectra were transformed into the rest frame using the redshift ($z=0.022189$). 
These processed spectra are adopted in the next analysis. 

\begin{figure}[ht!]
\centering
\includegraphics[angle=0,width=0.49\textwidth]{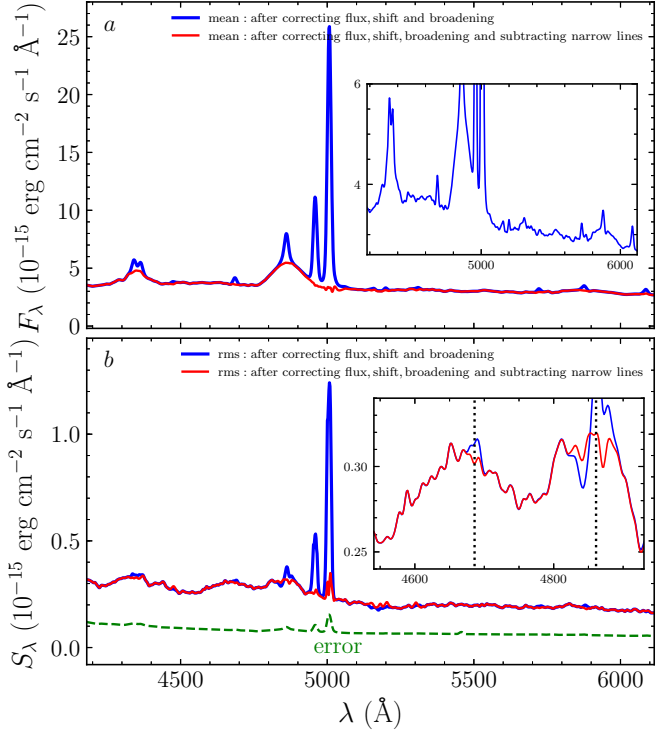}
\caption{\footnotesize
Panels show the mean ({\it a}) and rms ({\it b}) spectrum of Mrk79 in the rest frame, respectively. 
Blue lines are the original mean and rms spectrum calculated from processed spectra (see Section~\ref{sec_dp}). 
Red lines are the revised mean and rms spectrum constructed after subtracting narrow emission lines 
(including H$\gamma\lambda4340$, [O{\sc iii}]$\lambda4363$, He~{\sc i}$\lambda4471$, He~{\sc ii}$\lambda4686$, 
[Fe~{\sc vii}] $\lambda\lambda5158, 5178$, [N~{\sc i}] $\lambda5200$, [Ca~{\sc v}] $\lambda5310$, 
[Fe~{\sc vii}] $\lambda\lambda5721$, He~{\sc i}$\lambda5876$ and [Fe~{\sc vii}] $\lambda6086$) from each nightly spectrum. 
We only show the fitting region (4180\AA-6115\AA~in the rest frame, see Section~\ref{sec_fit}). 
}
\label{fig_mr}
\end{figure}

\begin{figure*}[ht!]
\centering
\includegraphics[angle=0,width=0.497\textwidth]{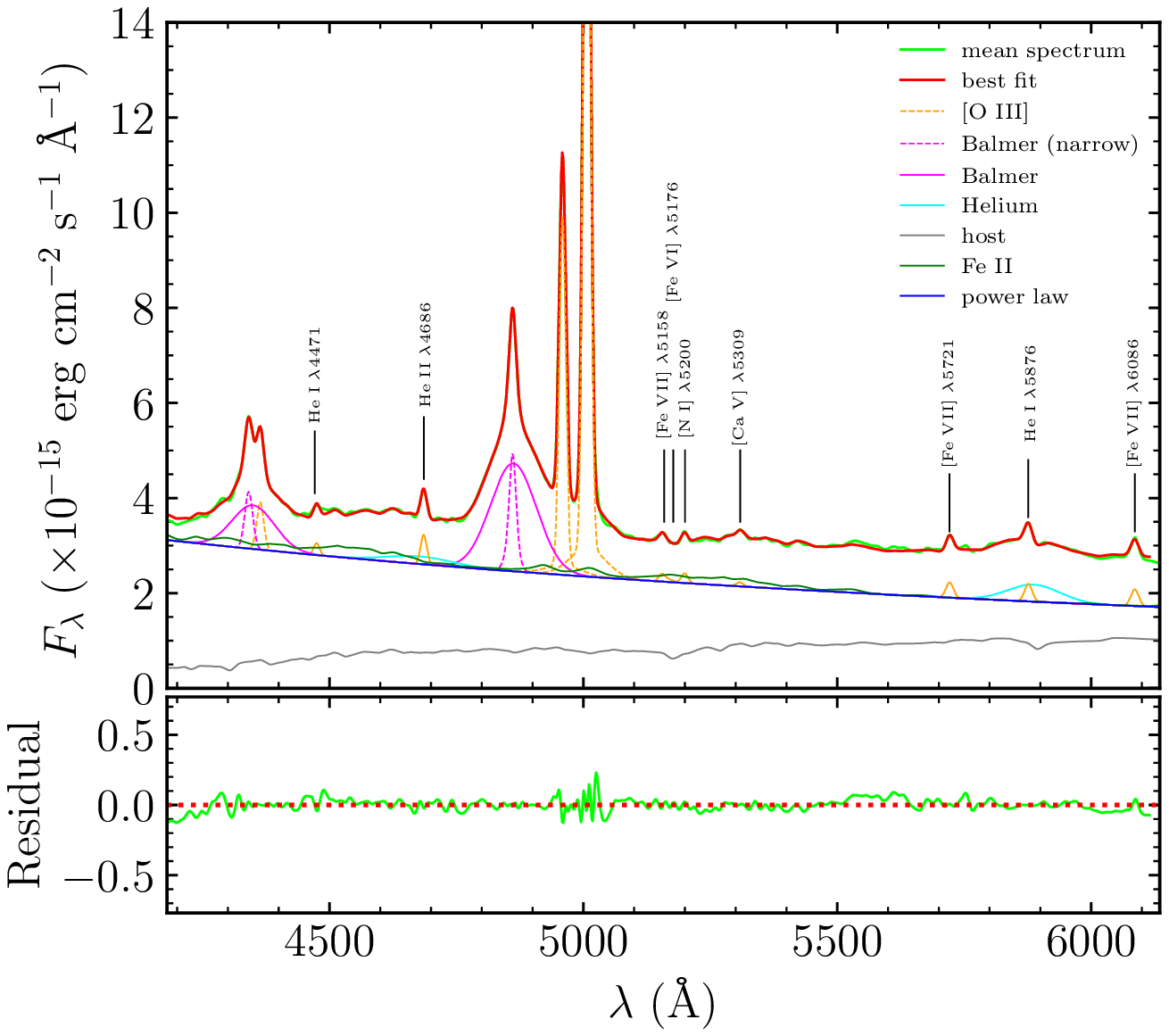}
\includegraphics[angle=0,width=0.497\textwidth]{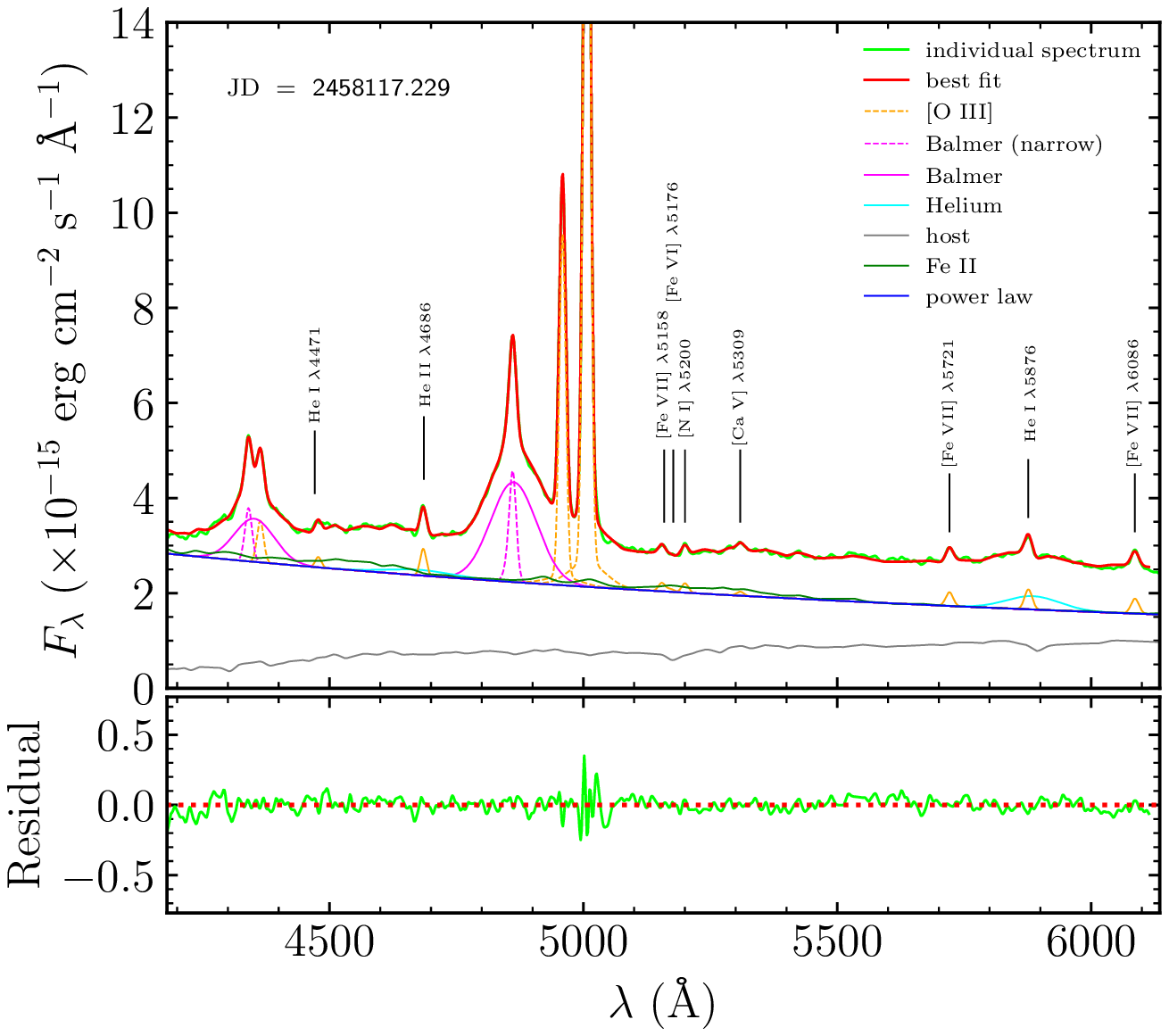}  
\caption{\footnotesize
Multi-components fitting of ({\it left}) the mean spectrum and ({\it right})
an individual spectrum of Mrk~79. 
The top traces show the processed spectrum (lime, see Section~\ref{sec_dp}) 
and the best-fit model (red), which is composed of AGN power-law continuum (blue), 
Fe~{\sc ii} multiplets  (green), host galaxy (gray), broad emission lines including 
H$\beta~\lambda4861$, H$\gamma~\lambda4340$ (magenta), He~{\sc ii}~$\lambda4686$ and He~{\sc i}~$\lambda5876$ (cyan), 
and several narrow emission lines (orange). 
The bottom trace shows the residuals (lime). 
}
\label{fig_fit}
\end{figure*}

\begin{figure}[ht!]
\centering
\includegraphics[angle=0,width=0.4\textwidth]{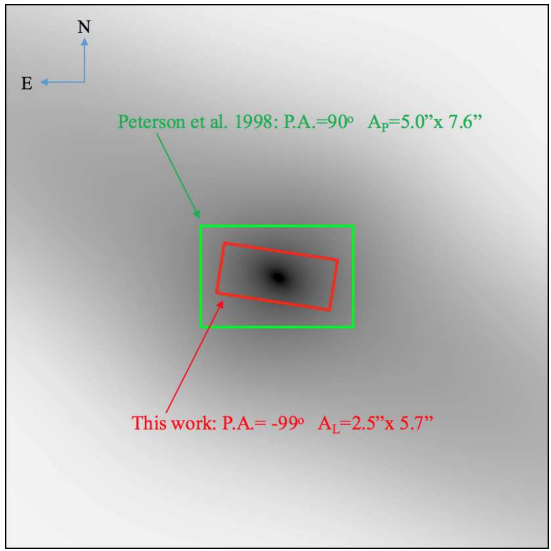}
\caption{\footnotesize
The best-fit model for host galaxy based on the two dimensional decomposition of the \textit{HST} image \citep{Kim2017}. 
Extracted apertures of spectrum are overlaid in red rectangle (adopted by this work) and green rectangle (adopted by \citealt{Peterson1998}). 
P.A. is position angle of slit. Image is displayed on logarithmic stretch with 25$^{\prime\prime}$$\times$25$^{\prime\prime}$. 
}
\label{fig_host}
\end{figure}

\section{Data Analysis and Results}
\label{sec_result}

\subsection{Mean and RMS Spectra}
\label{sec_meanrms}
The definition of mean spectrum is (\citealt{Peterson2004}) 
\begin{equation}
F_{\lambda}=\frac{1}{N}\sum_{i=1}^NF_i(\lambda) 
\label{eq_msp}
\end{equation}
and rms spectrum is 
\begin{equation}
S_{\lambda}=\left\{\frac{1}{N}\sum_{i=1}^N\left[F_i(\lambda)-\overline{F}(\lambda)\right]^2\right\}^{1/2}, 
\label{eq_rsp}
\end{equation}
where $F_i(\lambda)$ is the $i$th spectrum and $N$ is the total number of spectra obtained during the monitoring period. 
Using Equations~(\ref{eq_msp},\ref{eq_rsp}), we calculated the mean and rms spectrum of Mrk~79 
from the processed spectra and showed them in Figure~\ref{fig_mr} (in blue). 
Zoom-in of mean spectrum was inserted in Figure~\ref{fig_mr}{\it a},  which presents significant corona lines 
(such as [Fe~{\sc vii}]~$\lambda\lambda5158, 5178$, [Fe~{\sc vii}]~$\lambda5721$, 
[Fe~{\sc vii}]~$\lambda6086$ and [Ca~{\sc v}]~$\lambda5310$), 
narrow lines (e.g., He~{\sc i}~$\lambda4471$, He~{\sc i}~$\lambda5876$), 
and weak absorption-line features (e.g., see the red wing of He~{\sc i}~$\lambda5876$ narrow line). 
[O~{\sc iii}] emission lines should normally disappear in rms spectrum 
because the flux of [O~{\sc iii}] should not vary on the BLR reverberation timescale \citep{{Barth2016}}. 
However, Figure~\ref{fig_mr}{\it b} shows that [O~{\sc iii}]~$\lambda\lambda$4959,~5007 
emission lines dramatically remain in rms spectrum. 

In practice, 
two scenarios will result in the [O~{\sc iii}] remaining in the rms spectrum. 
(1) Residual [O~{\sc iii}] is caused by wavelength shift and broadening of emission line. 
(2) As considered in Section~\ref{sec_obs} (the third paragraph), residual [O~{\sc iii}] may 
be attributed to the apparent variation in flux of [O~{\sc iii}] caused 
by varying observing conditions because the narrow-line region (NLR) of Mrk~79 is slightly extended 
(\citealt{Schmitt2003a,Schmitt2003b,Ho2009}, also see Appendix~\ref{sec_a}). 
For the former, we have processed spectra of Mrk~79 strictly before calculating rms spectrum, 
therefore the residuals of [O~{\sc iii}] in rms spectrum do not caused by the shift and broadening of emission line. 
If the latter case holds, apparent variation in flux of [O~{\sc iii}] and host-galaxy should have similar behaviour 
(or be correlated) because NLR and the host galaxy are extended sources, 
we will examine this in Section~\ref{sec_lc}. 

Based on the mean and rms spectrum, 
we found that the broad He~{\sc ii}~$\lambda4686$ is very weak (see blue trace of Figure~\ref{fig_mr}{\it a}), 
but its flux shows significant variation (see blue trace of Figure~\ref{fig_mr}{\it b}) during the monitoring period. 
To determine the width of the broad He~{\sc ii}~$\lambda4686$ line from the rms spectrum, 
we run 200 Monte Carlo simulations similar to the method adopted by \cite{Grier2012}. 
We created 200 rms spectra from 200 randomly chosen subsets of the spectra, 
and obtained distributions of line width (FWHM) and positions of line core ($C$). 
The distributions give $\rm FWHM_{He~{\sc II}~\lambda4686}= 9621\pm812~km~s^{-1}$ 
and $C \rm _{He~{\sc II}~\lambda4686}=4678.6\pm3.9~\AA$. 
The latter corresponds to a blueshift of the broad He~{\sc ii}~$\lambda4686$ 
emission line with velocity of $\sim 450~\rm km~s~^{-1}$. 

\subsection{Spectral Fitting} 
\label{sec_fit}
To more accurately separate the broad emission features from each other, 
spectral fitting scheme (SFS) is widely used in the spectroscopic measurements of AGN 
(e.g, \citealt{Hu2008,Wang2009,Dong2008,Dong2011,Stern2012,Jin2012,Liu2018,Lu2019}). Especially, 
in the study of reverberation mapping field, 
SFS has been proven to be necessary to measure the light curves 
when broad emission line blended highly with each other \citep{Bian2010,Hu2015,Barth2013,Barth2015}. 
Beyond that, by modelling and removing the contamination of strong host galaxy 
which varies from night to night due to seeing and guiding variations, 
SFS can improve the measurement quality of light curves of continuum and broad emission lines \citep{Hu2015,Hu2016}. 

In order to decompose the spectra of Mrk 79 using SFS, we followed previous method described 
by \cite{Hu2015} with some changes described below. 
The fitting was performed in the rest wavelength range 4180~\AA$-$6115~\AA, which has no effect of the 
second-order (secondary spectrum) in our analysis because it's contamination occurs longer than  6250~\AA.  
The fitting components include: 
(1) a single power law ($f_{\lambda}\propto\lambda^{\alpha}$, $\alpha$ is the spectral index) for the AGN continuum. 
In practice, a single power-law is successfully used to fit AGN continuum over a broad region ($\sim$4150~\AA$-$6200~\AA; \citealt{Hu2015}); 
(2) the starlight from the host galaxy modelled by the template with 11 Gyr age and metallicity Z = 0.05 from \cite{Bruzual2003};  
(3) Fe~{\sc ii} multiplets modelled by Fe~{\sc ii} template from \cite{Boroson1992} convolving with a Gaussian function; 
(4) four single Gaussians for the broad-emission lines including H$\beta~\lambda4861$, 
H$\gamma~\lambda4340$, He~{\sc ii}~$\lambda4686$, and He~{\sc i}~$\lambda5876$, respectively; 
(5) three double Gaussians for the [O~{\sc III}] doublets $\lambda5007/\lambda$4959 
and H$\beta~\lambda4861$ narrow line; 
(6) a set of several single Gaussians with the same velocity width and shift for narrow emission lines 
including H$\gamma~\lambda4340$, [O~{\sc iii}]~$\lambda4363$, He~{\sc i}~$\lambda4471$, 
He~{\sc ii}~$\lambda4686$, He~{\sc i}~$\lambda5876$, [N~{\sc i}]~$\lambda5200$ and several coronal lines. 
Following \cite{Hu2015}, we fitted the above models simultaneously to the spectra of Mrk~79 in the fitting region. 
The processed spectra (see Section~\ref{sec_dp}) were fitted in two steps. 
We fitted the mean spectrum at first. During the fitting, 
the flux ratio of [O~{\sc iii}] doublets was fixed to the theoretical value of 3. 
The shift and line width of the broad He~{\sc ii}~$\lambda4686$ emission line were 
fixed to the best value measured in Section~\ref{sec_dp} 
because it's too weak to restrict the model. 
The rest of model parameters were allowed to vary. 
Then in the fitting of individual spectrum, 
we fixed the spectral index, 
the flux ratios of the narrow emission lines relative to [O~{\sc iii}]~$\lambda5007$  
to the corresponding values given by the best fit of the mean spectra. 
The spectra of Mrk 79 show weak features of Fe~{\sc ii} multiplets, 
so we also fixed its width to the value fitted in the mean spectrum. 
In practice, 
this operation is reasonable since we have corrected the broadening of emission lines (see Section~\ref{sec_dp}). 

Using the fitting results, we calculated the revised mean and rms spectra 
after subtracting of the narrow emission lines, 
and over-plotted along with the original mean and rms spectra in Figure~\ref{fig_mr}. 
Comparing revised mean spectrum (in red) with original mean spectrum (in blue), 
we found that the narrow emission lines are well fitted and subtracted. 
From Figure~\ref{fig_mr}{\it b}, we found that the residuals of [O~{\sc III}]~$\lambda\lambda$4959,~5007 
almost disappear in the revised rms spectrum, 
minor residuals are comparable with mean errors (in dashed green-line). 

\subsection{Host galaxy}
Mrk~79 was observed by the \textit{HST} (\textit{Hubble Space Telescope}) ACS/HRC 
(Advanced Camera for Surveys/High resolution channel) with F550M filter. 
Two-dimensional surface brightness decomposition of Mrk~79 was performed by 
\cite{Bentz2009a} and \cite{Kim2017} using the code {\tt GALFIT} \citep{Peng2002,Kim2008}. 
\cite{Kim2017} recently analysed high-resolution \textit{HST} images of 235 low-redshift Type 1 AGNs 
to study the structures of the host galaxy. 
We adopted the best-fit model for the host galaxy of Mrk~79 from \cite{Kim2017}, which is shown in Figure~\ref{fig_host}. 
\cite{Bentz2009a} also analysed \textit{HST} images of 35 RM AGNs to measure 
the contribution of host light to the total luminosity at 5100~\AA. 
They concluded the flux of host light at 5100~\AA~for Mrk~79 is 
$F_{\rm host,A_{P}}[1+z]=1.42\times10^{-15}$~erg~s$^{-1}$~cm$^{-2}$~\AA$^{-1}$ within an aperture of 
5.0$^{\prime\prime}$$\times$7.6$^{\prime\prime}$ ($\rm A_{P}$; lime rectangle). 
In this work, the extraction aperture of spectrum is 2.5$^{\prime\prime}$$\times$5.7$^{\prime\prime}$ 
with a position angle $-99^{\circ}$ ($\rm A_{L}$; red rectangle in Figure~\ref{fig_host}). 
Integrating the photons in the extraction apertures of 2.5$^{\prime\prime}$$\times$5.7$^{\prime\prime}$ 
and 5.0$^{\prime\prime}$$\times$7.6$^{\prime\prime}$, respectively, 
we obtained the ratio of total photons of the host galaxy in two apertures $\rm A_{L}/A_{P}=0.60$. 
Using this ratio and the host-galaxy flux $1.42\times10^{-15}$~erg~s$^{-1}$~cm$^{-2}$~\AA$^{-1}$, 
we estimated the host-galaxy flux in our adopted extraction aperture for Mrk~79, 
which yields $F_{\rm host,A_{L}}[1+z]=0.85\times10^{-15}$~erg~s$^{-1}$~cm$^{-2}$~\AA$^{-1}$. 
The fitting of mean spectrum yields an average host-galaxy flux of 
$F_{\rm host}[1+z]=(0.80\pm0.09)\times10^{-15}$~erg~s$^{-1}$~cm$^{-2}$~\AA$^{-1}$, 
where uncertainty 0.09 is the standard deviation estimated from the fitted host-galaxy components (Section~\ref{sec_fit}). 
This flux is consistent with the above estimate from \textit{HST} image 
because measuring the flux contribution from the image has an uncertainty of $\sim10\%$ \citep{Bentz2013}. 
This consistency indirectly shows that the spectral decomposition of Mrk~79 is robust and the host galaxy of Mrk~79 
is well fit with a bulge, a bar, and a disk (see \citealt{Kim2017}). 

\begin{figure}[ht!]
\includegraphics[angle=0,width=0.5\textwidth]{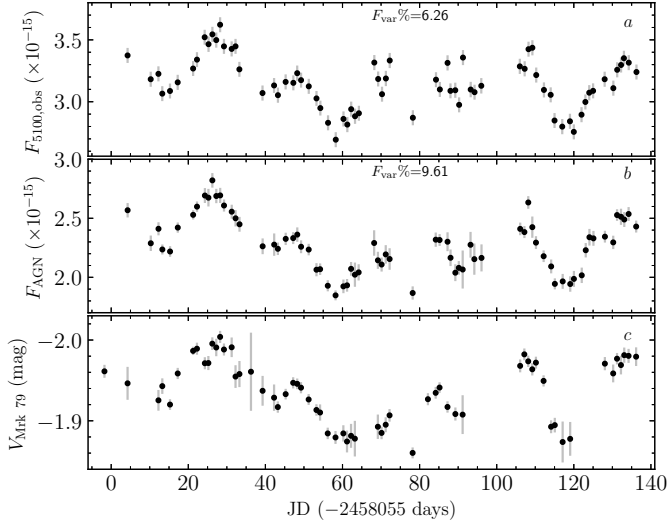}
\caption{\footnotesize 
Comparison of light curves. 
Panels ({\it a-c}) are light curves of the observed continuum at 5100\AA~(from the processed spectra), 
AGN continuum (from the featureless power law), 
and Mrk~79 photometry (see Section~\ref{sec_lc}), respectively. 
}
\label{fig_lc}
\end{figure}

\subsection{Light Curves}
\label{sec_lc}
The light curves of AGN~continuum at 5100~\AA~($F_{\rm AGN}$) and broad emission lines 
($F_{\rm H\beta~\lambda4861}$, $F_{\rm H\gamma~\lambda4340}$, 
$F_{\rm He~{\sc II}~\lambda4686}$, and $F_{\rm He~{\sc I}~\lambda5876}$) 
were generated from the best-fit model (the featureless power law and the broad line components). 
Table~\ref{tab_lcccf} provides the data of these light curves along with the photometric light curve of Mrk~79. 
We also measured the fluxes of the host galaxy ($F_{\rm gal}$) 
and [O~{\sc iii}]~$\lambda5007$ ($F_{\rm [O~{\sc III}]}$) from the best-fit model, 
and directly measured the light curve of continuum at 5100~\AA~($F_{\rm 5100,obs}$, 
which is contaminated by the host galaxy) 
from the above processed spectra (see Section~\ref{sec_dp}). 
We calculated the variability amplitude of the light curve by (see \citealt{Rodriguez-Pascual1997}) 
\begin{equation}
F_{\rm var}=\frac{\left(\sigma^2-\Delta^2\right)^{1/2}}{\langle F\rangle} \, 
\label{eq_fvar}
\end{equation}
and its uncertainty (\citealt{Edelson2002}) 
\begin{equation}
\sigma_{_{F_{\rm var}}} = \frac {1} {F_{\rm var}} \left(\frac {1}{2 N}\right)^{1/2} \frac {\sigma^2}{\langle F\rangle} \, ,
\end{equation}
where $\langle F\rangle=N^{-1}\sum_{i=1}^NF_i$ is the average flux, $F_i$ is the flux 
of the $i$-th observation of the light curve, $N$ is the total number of observations, 
$\sigma^2=\sum_{i=1}^N\left(F_i-\langle F\rangle\right)^2/(N-1)$, 
$\Delta^2=\sum_{i=1}^N\Delta_i^2/N$, 
and $\Delta_i$ is the uncertainty of $F_i$. Table~\ref{tab_lcst} lists the statistics 
of the light curve. 

We compared the light curves of $F_{\rm AGN}$ and $F_{\rm 5100,obs}$ (Figure~\ref{fig_lc}$a$ and~\ref{fig_lc}$b$), 
and found that the light curve of $F_{\rm AGN}$ has the larger variability amplitude ($F_{\rm var}\%=9.61$) 
than the light curve of $F_{\rm 5100,obs}$ ($F_{\rm var}\%=6.26$) . 
Which shows that the light curve of $F_{\rm AGN}$ has improved after removing the contamination of the host galaxy. 
Both light curves measured from the spectra are consistent with the photometric light curve of Mrk~79 (Figure~\ref{fig_lc}$c$). 
This consistency shows that the stable comparison star can provide precise flux calibration of spectra. 

We also checked the fluxes of the host galaxy and the [O~{\sc iii}]~$\lambda5007$ (measured from the best-fit components) 
in Figure~\ref{fig_elc}$a$~and~\ref{fig_elc}$b$, found that the both fluxes have similar apparent variation 
in the time domain, and found that the host-galaxy fluxes (11\%) are more scatter than [O~{\sc iii}]~$\lambda5007$ (5\%). 
Similar phenomena are found in Mrk~382 by \cite{Hu2015}. 
These measurement results are consistent with the considerations of spectroscopic observation 
in Section~\ref{sec_obs} (the third paragraph). 
That is monitoring the spectra of Mrk~79 in a relatively narrow slit ($2.5^{\prime\prime}$) contributes to increase 
the spectral resolution, but varying observing conditions cause the apparent variation in flux of the 
slightly extended components observed in the narrow slit (including the [O III] emission region and the host galaxy). 
The apparent variation in flux of the [O~{\sc iii}]~$\lambda5007$ leads to the [O~{\sc iii}] remaining 
in the rms spectrum, which supports the second scenario mentioned in Section~\ref{sec_meanrms}. 

In order to qualitatively study the apparent variation in flux of the host galaxy and the [O III] λ5007, 
we have insight into the details of spectroscopy and flux calibration in Appendix~\ref{sec_a}. 
Briefly, seeing is a major factor in responding to varying observing conditions, 
which could change from one exposure to the next. 
Two point sources (the comparison star and AGN) kept in a line parallel to the narrow slit ($2.5^{\prime\prime}$), 
the fractions of light loss due to varying seeing are identical (Figure~\ref{fig_sch}), 
that is the stable comparison star can provide precise flux calibration of spectra (see figure~\ref{fig_lc} and accompanying statements). 
However, the slightly or very extended components in the same slit, 
the fractions of light loss due to varying seeing are less than the point source (Figure~\ref{fig_sch} of Appendix~\ref{sec_a}). 
In practice, above analysis is consistent with seeing-induced aperture effects addressed by \cite{Peterson1995}. 
Consequently, 
the calibrated fluxes of the slightly extended components should correlate with varying seeing; 
The host-galaxy fluxes 
should be more scatter than the [O~{\sc iii}]'s fluxes with varying seeing, 
because the intrinsic size of the host galaxy is larger than the [O~{\sc iii}] emission region.  
The elaborate analysis is provided in Appendix~\ref{sec_a}. 
Ultimately, our measurement results in fluxes of the slightly extended components (i.e. Figure~\ref{fig_elc} and \ref{fig_seecor}) 
from the spectral fitting productions are consistent with above analysis (details refer to Appendix~\ref{sec_a}), 
this consistency demonstrates that the spectral fitting scheme performs correct decomposition of multi-components.  

\begin{deluxetable*}{cccccccc}
\tablecolumns{12}
\tabletypesize{\scriptsize}
\tablewidth{0pt}
\tablecaption{Light curves of continuum at 5100~\AA~and broad emission lines for Mrk~79\label{tab_lcccf}}
\tablehead{
\colhead{JD$-$2450000}   &
\colhead{$V$-band~(mag)}                &
\colhead{JD$-$2450000}   &
\colhead{$F_{\rm AGN}$}                &
\colhead{$F_{\rm H\gamma}$}        &
\colhead{$F_{\rm He~{\sc II}}$}       &
\colhead{$F_{\rm H\beta}$}             &
\colhead{$F_{\rm He~{\sc I}}$}         
}
\startdata
8053.25 &$ -1.96 \pm 0.01 $& 8059.26 &$ 2.57 \pm 0.06 $&$ 1.03 \pm 0.05 $&$ 3.26 \pm 0.31 $&$ 2.65 \pm 0.04 $&$ 4.71 \pm 0.17 $\\
8059.25 &$ -1.95 \pm 0.02 $& 8065.27 &$ 2.29 \pm 0.07 $&$ 1.03 \pm 0.04 $&$ 2.07 \pm 0.31 $&$ 2.60 \pm 0.04 $&$ 4.73 \pm 0.17 $\\
8067.26 &$ -1.93 \pm 0.01 $& 8067.27 &$ 2.41 \pm 0.06 $&$ 0.98 \pm 0.04 $&$ 2.49 \pm 0.29 $&$ 2.64 \pm 0.04 $&$ 4.87 \pm 0.16 $\\
8068.25 &$ -1.94 \pm 0.01 $& 8068.27 &$ 2.24 \pm 0.04 $&$ 0.96 \pm 0.04 $&$ 1.64 \pm 0.27 $&$ 2.56 \pm 0.03 $&$ 4.71 \pm 0.12 $\\
8070.25 &$ -1.92 \pm 0.01 $& 8070.27 &$ 2.22 \pm 0.05 $&$ 0.95 \pm 0.04 $&$ 0.95 \pm 0.27 $&$ 2.54 \pm 0.03 $&$ 4.71 \pm 0.14 $
\enddata
\tablecomments{\footnotesize
$V$-band is photometric data (instrumental magnitude), 
$F_{\rm AGN}$ is AGN continuum at 5100~\AA~in units of ${\rm 10^{-15}~erg~s^{-1}~cm^{-2}~\AA^{-1}}$, 
$F_{\rm H\gamma}$ and $F_{\rm H\beta}$ are the fluxes of H$\gamma~\lambda4340$ and H$\beta~\lambda4861$ 
in units of ${\rm 10^{-13}~erg~s^{-1}~cm^{-2}}$, 
$F_{\rm He~{\sc II}}$ and $F_{\rm He~{\sc I}}$ are the fluxes of He~{\sc ii}~$\lambda4686$ and He~{\sc i}~$\lambda5876$ 
in units of ${\rm 10^{-14}~erg~s^{-1}~cm^{-2}}$. 
This table is available in its entirety in machine-readable form. 
}
\end{deluxetable*}

\begin{deluxetable*}{lcccccc}
\tablecolumns{7}
\tabletypesize{\scriptsize}
\tablewidth{0pt}
\tablecaption{Statistics of light curve for Mrk~79 in this campaign\label{tab_lcst}}
\tablehead{
\colhead{Time Series}                    &
\colhead{$F_{\rm 5100,obs}[1+z]$} &
\colhead{$F_{\rm AGN}[1+z]$} &
\colhead{$F_{\rm H\gamma}[1+z]$} &
\colhead{$F_{\rm He~{\sc II}}[1+z]$} &
\colhead{$F_{\rm H\beta}[1+z]$} &
\colhead{$F_{\rm He~{\sc I}}[1+z]$} 
}
\startdata
Mean flux                 & $3.22\pm0.21$  &   2.33$\pm$0.23    &  1.00$\pm$0.10 &  2.63$\pm$1.21   &  2.61$\pm$0.14  &  4.64$\pm$0.42  \\
$F_{\rm var}$ (\%)   & $6.26\pm0.57$ &   9.61$\pm$0.88 &  8.53$\pm$0.92  &  45.37$\pm$ 4.06 &  5.33$\pm$0.48  &  8.28$\pm$0.85  
\enddata
\tablecomments{\footnotesize
The mean flux $F_{\rm 5100,obs}[1+z]$ in units of ${\rm 10^{-15}~erg~s^{-1}~cm^{-2}~\AA^{-1}}$, 
the mean flux of other time series have same units with Table~\ref{tab_lcccf}. 
}
\end{deluxetable*}

\begin{deluxetable*}{ccccccc}
\tablecolumns{2}
\tabletypesize{\scriptsize}
\tablewidth{4pt}
\tablecaption{Summary of previous RM results of Mrk~79\label{tab_opl}}
\tablehead{
\colhead{Epoch (JD: days)}                           &
\colhead{$F_{\rm obs}[1+z]$}                  &
\colhead{$F_{\rm host}[1+z]$}              &
\colhead{$F_{\rm AGN}[1+z]$}             &
\colhead{$L_{5100}$~(${\rm erg~s^{-1}}$)}                  &
\colhead{${\rm H\beta~Lags~(days)}$}                  &
\colhead{Ref.}      
}
\startdata
2447838$-$2448044 & 6.96$\pm$0.15 & $1.42\pm0.07$ & 5.54$\pm$0.18 & $(3.45\pm0.12)\times10^{43}$ & $  9.0^{+8.3}_{-7.8}$ & (1,2) \\
2448193$-$2448393 & 8.49$\pm$0.16 & $1.42\pm0.07$ & 7.07$\pm$0.19 & $(4.41\pm0.11)\times10^{43}$ & $16.1^{+6.6}_{-6.6}$ & (1,2) \\
2448905$-$2449135 & 7.40$\pm$0.16 & $1.42\pm0.07$ & 5.98$\pm$0.19 & $(3.73\pm0.12)\times10^{43}$ & $16.0^{+6.4}_{-5.8}$ & (1,2) \\
2458059$-$2458192 & 3.22$\pm$0.22 & $0.80\pm0.09$ & 2.33$\pm$0.23 & $(1.45\pm0.14)\times10^{43}$ & $3.49^{+0.62}_{-0.60}$ & (3)         
\enddata
\tablecomments{\footnotesize
{\bf References}: (1) \cite{Peterson1998}; (2) \cite{Bentz2013}; (3) This work, H$\beta$ lag from Table~\ref{tab_rm}. \\
The units of $F_{\rm obs}[1+z]$, $F_{\rm host}[1+z]$, and $F_{\rm AGN}[1+z]$ are ${\rm 10^{-15}~erg~s^{-1}~cm^{-2}~\AA^{-1}}$. \\
}
\end{deluxetable*}

\begin{deluxetable*}{lcccccccccccc}
  \tablecolumns{14}
  \tabletypesize{\scriptsize}
  \setlength{\tabcolsep}{3pt}
  \tablewidth{4pt}
  \tablecaption{RM measurements of Mrk~79 from the present campaign\label{tab_rm}}
  \tablehead{
  \colhead{} &
  \multicolumn{3}{c}{Line vs.~$F_{\rm AGN}$} & 
  \colhead{} &
  \multicolumn{2}{c}{Mean spectra} & 
  \colhead{} &
  \multicolumn{2}{c}{rms spectra} & 
  \colhead{}  \\ \cline{2-4} \cline{6-7} \cline{9-10} 
   \colhead{Lines}&
  \colhead{$\tau_{\rm cent}$}&
   \colhead{$\tau_{\rm peak}$}&
  \colhead{$r_{\rm max}$}&
  \colhead{} &
  \colhead{FWHM~(km~s$^{-1}$)}&
  \colhead{$\sigma_{\rm line}$~(km~s$^{-1}$)}&
  \colhead{} &
  \colhead{FWHM~(km~s$^{-1}$)}&
  \colhead{$\sigma_{\rm line}$~(km~s$^{-1}$)}&
  \colhead{VP ($\times10^{7} M_{\odot}$)} &
   \colhead{VP ($\times10^{7} M_{\odot}$)} \\
  \colhead{(1)} &
  \colhead{(2)} &
  \colhead{(3)} &
  \colhead{(4)} &
  \colhead{} &
  \colhead{(5)} &
  \colhead{(6)} &
  \colhead{} &
  \colhead{(7)} &
  \colhead{(8)} &
  \colhead{(9)} &
  \colhead{(10)}
}
\startdata
H$\gamma$ & $2.43^{+1.55}_{0.88}$ &$2.32^{+1.00}_{-1.17}$ &  0.79  && 6730$\pm$312    & 2739$\pm$216 && 10757$\pm$377 & 3679$\pm$137    & $2.15^{+1.38}_{-0.80}$ & $0.64^{+0.41}_{-0.24}$ \\   
He~{\sc ii}    & $-0.05^{+0.50}_{-0.40}$ &$0.21^{+0.28}_{-0.77}$ &  0.86 &&  ---                      & ---                      &&  9585$\pm$815 & 4003$\pm$121     &  --- &  --- \\   
H$\beta$     & $3.49^{+0.62}_{-0.60}$ &$2.36^{+2.03}_{-0.49}$ &   0.83  &&  6539$\pm$154   & 2660$\pm$136 && 8431$\pm$621  & 3458$\pm$132     & $2.91^{+0.54}_{-0.52}$ & $0.82^{+0.16}_{-0.15}$\\   
He~{\sc i}    & $4.39^{+1.26}_{-1.12}$ &$4.43^{+1.73}_{-1.80}$ &   0.69  &&  6124$\pm$223  &  2470$\pm$156 && 8182$\pm$598  & 3345$\pm$134      & $3.21^{+0.95}_{-0.85}$  & $0.96^{+0.28}_{-0.25}$    
\enddata  
\tablecomments{\footnotesize
The Lags ($\tau_{\rm cent}$ and $\tau_{\rm peak}$) are in the rest frame. 
The VPs of Col. (9) are calculated using the measurements of Col. (2) and Col. (5), 
and Col. (10) are calculated using the measurements of Col. (2) and Col. (8).
}
\end{deluxetable*}

\subsection{Optical luminosity}
\label{sec_opl}
Using the light curve of AGN continuum at 5100~\AA~generated from the best-fit power law, we obtained the mean flux of 
$F_{\rm AGN}[1+z]=(2.33\pm0.23)~\times~{\rm 10^{-15}~erg~s^{-1}~cm^{-2}~\AA^{-1}}$, 
which corresponds the monochrome luminosity of $L_{5100}=(1.45\pm0.14)\times10^{43}~{\rm erg~s^{-1}}$ in the present epoch. 
Before this RM campaign, three-season RM campaigns for Mrk79 were finished by \cite{Peterson1998}, 
the data were updated and published in a series works (e.g., \citealt{Peterson2004,Zu2011,Bentz2013}). 
Primary parameters along with values are summarised in Table~\ref{tab_opl}, 
where the values of H$\beta$ lags quoted in Table~\ref{tab_opl} are compiled from latest result of \cite{Bentz2013}. 
Comparing these fluxes ($F_{\rm AGN}[1+z]$) for the host galaxy correction, 
we found that Mrk 79 is seeming to come out the faint state during the monitoring period. 
The mean flux approximates a magnitude fainter than previous record holder. 
We checked the historical data \citep{Peterson1998}, 
and found that the highest luminosity state of Mrk 79 appeared at JD$\sim$2448400 days. 
Similar to the famous NGC 5548 \citep{Lu2016}, the huge variation of AGN continuum benefit us to 
(1) investigate the variation of the BLR similar to NGC~5548 \citep{Lu2016}; 
(2) construct the Radius$-$Luminosity relationship of Mrk~79, 
similarly see $R_{\rm BLR}-L_{\rm 5100}$ relationship of NGC~5548 \citep{Lu2016,Pei2017}.

\subsection{Line profile measurements}
\label{sec_profile}
The broad emission lines including H$\gamma~\lambda4340$, H$\beta~\lambda4861$, 
and He~{\sc i}~$\lambda5876$ were covered in our spectral fitting window.  
Hence, we measured the velocity widths of these broad emission lines from the best-fit model. 
Using Equation~\ref{eq_msp}~and~\ref{eq_rsp}, we obtained the mean and rms spectrum of these 
broad lines at first. 
They velocity widths (FWHM \& $\sigma_{\rm line}$) are used as the optimum values. 
As adopted in Section~\ref{sec_meanrms}, to estimate the corresponding errors,  
we generated 200 mean and rms spectra (realizations) from 200 randomly chosen subsets of the spectra, 
and measured all velocity widths from simulated realizations. 
Then we used the standard deviations as errors of the optimum values. 
\cite{Whittle1992} obtained [O~{\sc iii}]~$\lambda5007$ width (FWHM$=350~{\rm km~s^{-1}}$) of Mrk~79 from 
high resolution spectrum. Comparing this width to those from our spectra (FWHM$=900~{\rm km~s^{-1}}$), 
we obtained a broadening of $829~{\rm km~s^{-1}}$. After correcting the broadening, we 
listed the widths of the broad lines in Table~\ref{tab_rm}. 
In spectral fitting, the velocity width of broad He~{\sc ii}~$\lambda4686$ line was fixed to the value 
estimated from rms spectrum (see Section~\ref{sec_meanrms}), 
therefore we only give its velocity width from rms spectrum in Table~\ref{tab_rm}. 

\begin{figure*}[ht!]
\centering
\includegraphics[angle=0,width=0.9\textwidth]{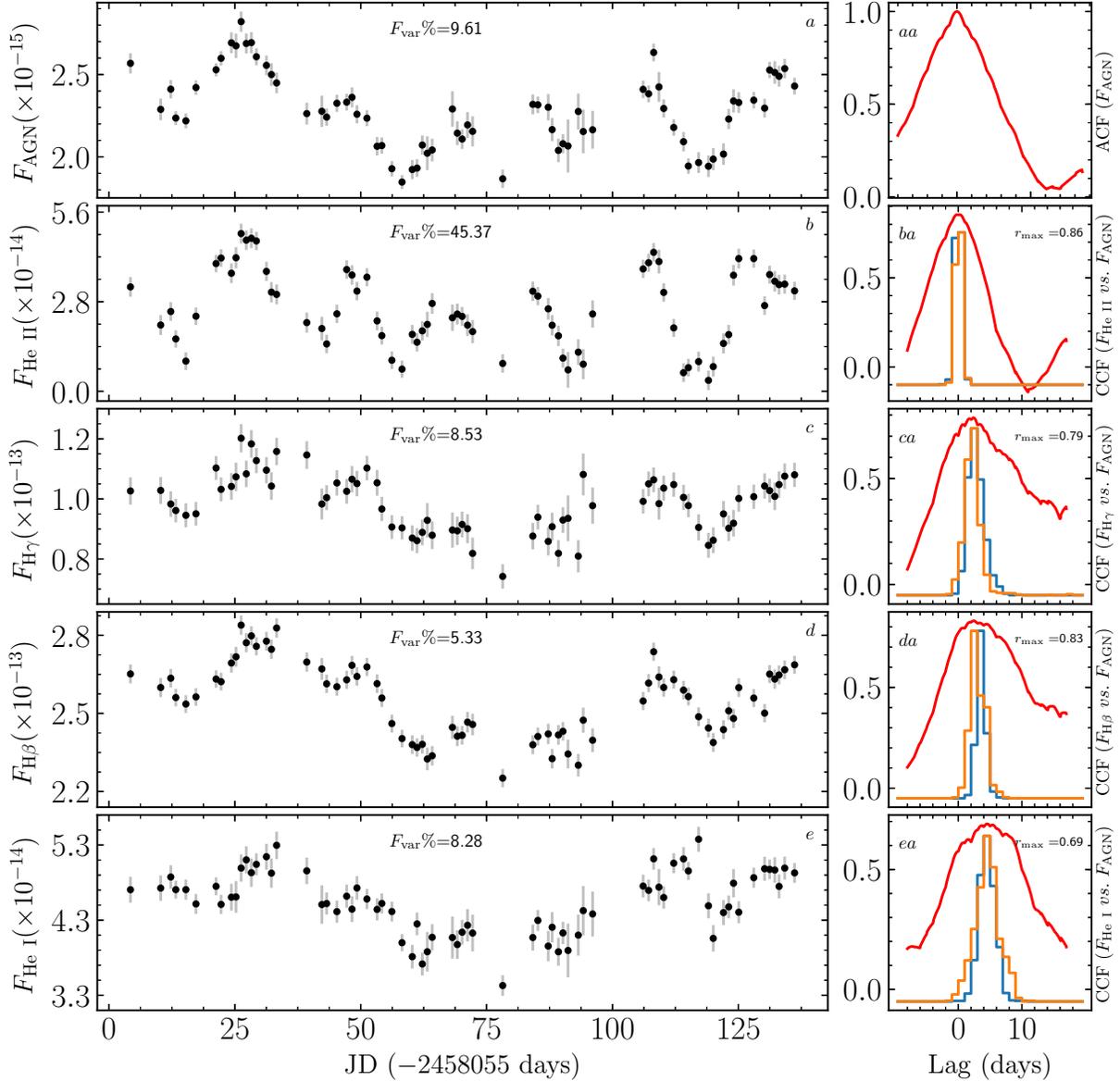}
\caption{\footnotesize
Light curves and the results of cross correlation analysis. 
The left panels ({\it a-e}) are the light curves of AGN continuum at 5100~\AA~and the broad emission lines 
calculated from the best-fit component. 
The right panels ({\it aa-ea}) correspond to the ACF of continuum and the CCF between the light curves 
of broad emission lines ({\it b-e}) and the continuum variation ({\it a}), respectively. 
We noted the variability amplitude $F_{\rm var}\%$ in panel of light curves, and noted the maximum 
cross-correlation coefficient ($r_{\max}$) in panel of CCF. 
The units of $F_{\rm AGN}$ and emission lines (including Helium and Balmer)
are ${\rm erg~s^{-1}~cm^{-2}~\AA^{-1}}$ and ${\rm erg~s^{-1}~cm^{-2}}$, respectively. 
}
\label{fig_mlc}
\end{figure*}

\subsection{Lags of the broad emission lines}
\label{sec_lag}
We measured the reverberation lags of the broad emission lines (H$\beta~\lambda4861$, 
H$\gamma~\lambda4340$, He~{\sc ii}~$\lambda4686$, and He~{\sc i}~$\lambda5876$) 
with respect to the continuum variation ($F_{\rm AGN}$), using the standard interpolation 
cross-correlation function (ICCF) method \citep{Gaskell1986,Gaskell1987,White1994}. 
The reverberation lags are usually measured from peak ($\tau_{\rm peak}$) and centroid ($\tau_{\rm cent}$ ) of the ICCF, 
where $\tau_{\rm peak}$ corresponds the maximum correlation coefficient $r_{\rm max}$, 
and $\tau_{\rm cent}$ is measured around the peak above a typical value ($r\ge0.8~r_{\rm max}$). 
The uncertainties of $\tau_{\rm peak}$ and $\tau_{\rm cent}$ were obtained using the Monte Carlo 
``flux randomization and random subset sampling'' method described by \citet{Peterson1998} 
and \citet{Peterson2004}. The Monte Carlo simulations were run with 5000 realizations, 
and the cross-correlation peak and centroid distribution (CCPD and CCCD) 
were created from the generated samples. The uncertainties of $\tau_{\rm peak}$ 
and $\tau_{\rm cent}$ were then calculated from the CCPD and CCCD, respectively, 
with a 68.3\% confidence level ($1\sigma$). 

Table~\ref{tab_rm} lists the lags of the broad emission lines including $\tau_{\rm cent}$, $\tau_{\rm peak}$, 
and the maximum cross correlation coefficients ($r_{\max}$). 
In the low luminosity state of Mrk~79 (see Section~\ref{sec_opl}), 
the lag of H$\beta~\lambda4861$ relative to the continuum variation is 
significant shorter than the results of early RM campaigns  (see Table~\ref{tab_opl}). 
This is similar to the finding in better-observed NGC~5548 that the BLR size shortens with luminosity decreasing 
(e.g., \citealt{Peterson1999,Peterson2002,Bentz2007,DeRosa2018,Lu2016,Kriss2019ApJ}). 
In addition, the lag of He~{\sc ii}~$\lambda4686$ relative to the continuum variation approximates zero, 
which is consistent with other objects (e.g., Mrk~1511, see \citealt{Barth2013}). 

\begin{figure*}[ht!]
\centering
\includegraphics[angle=0,width=0.46\textwidth]{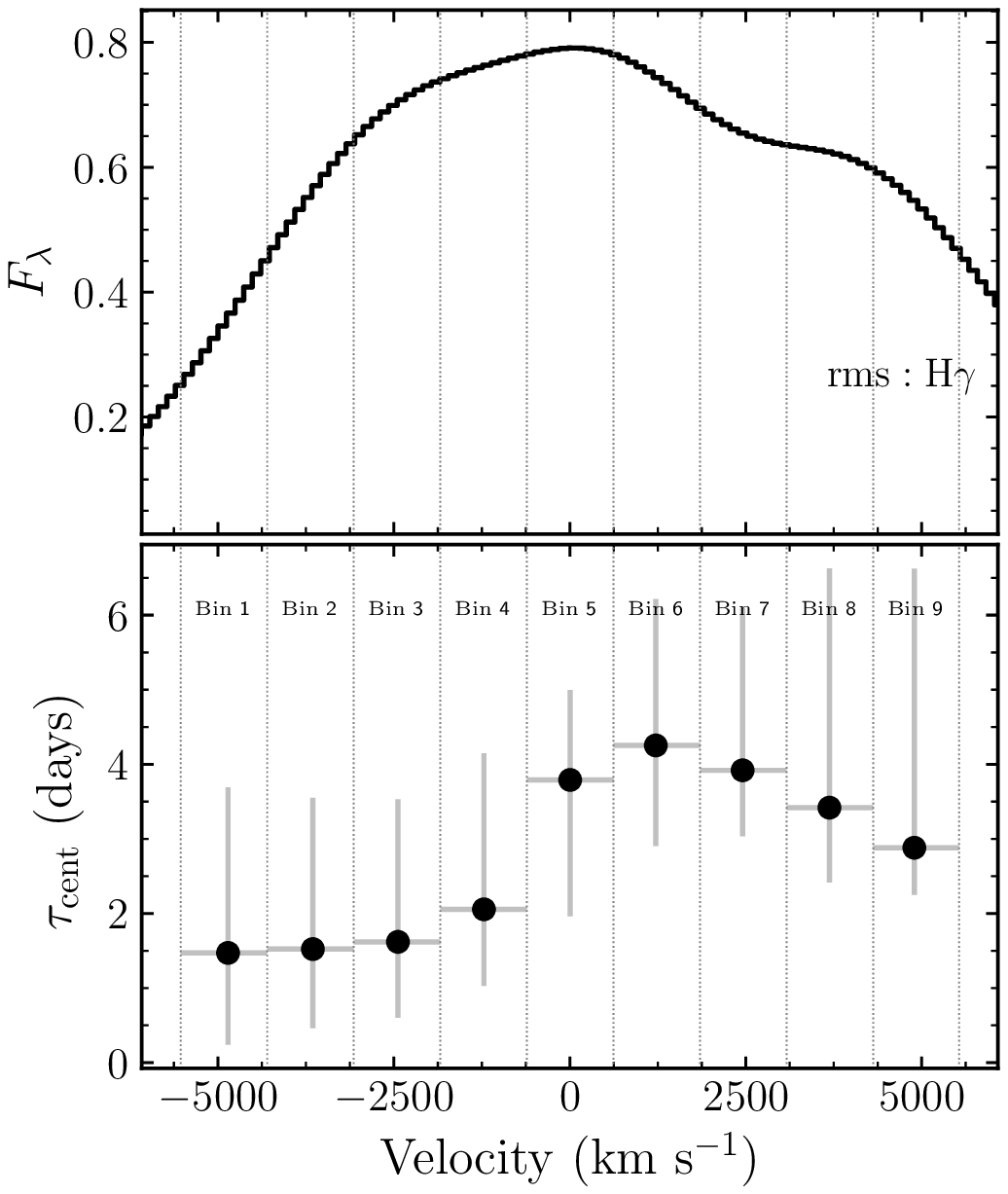}
\includegraphics[angle=0,width=0.45\textwidth]{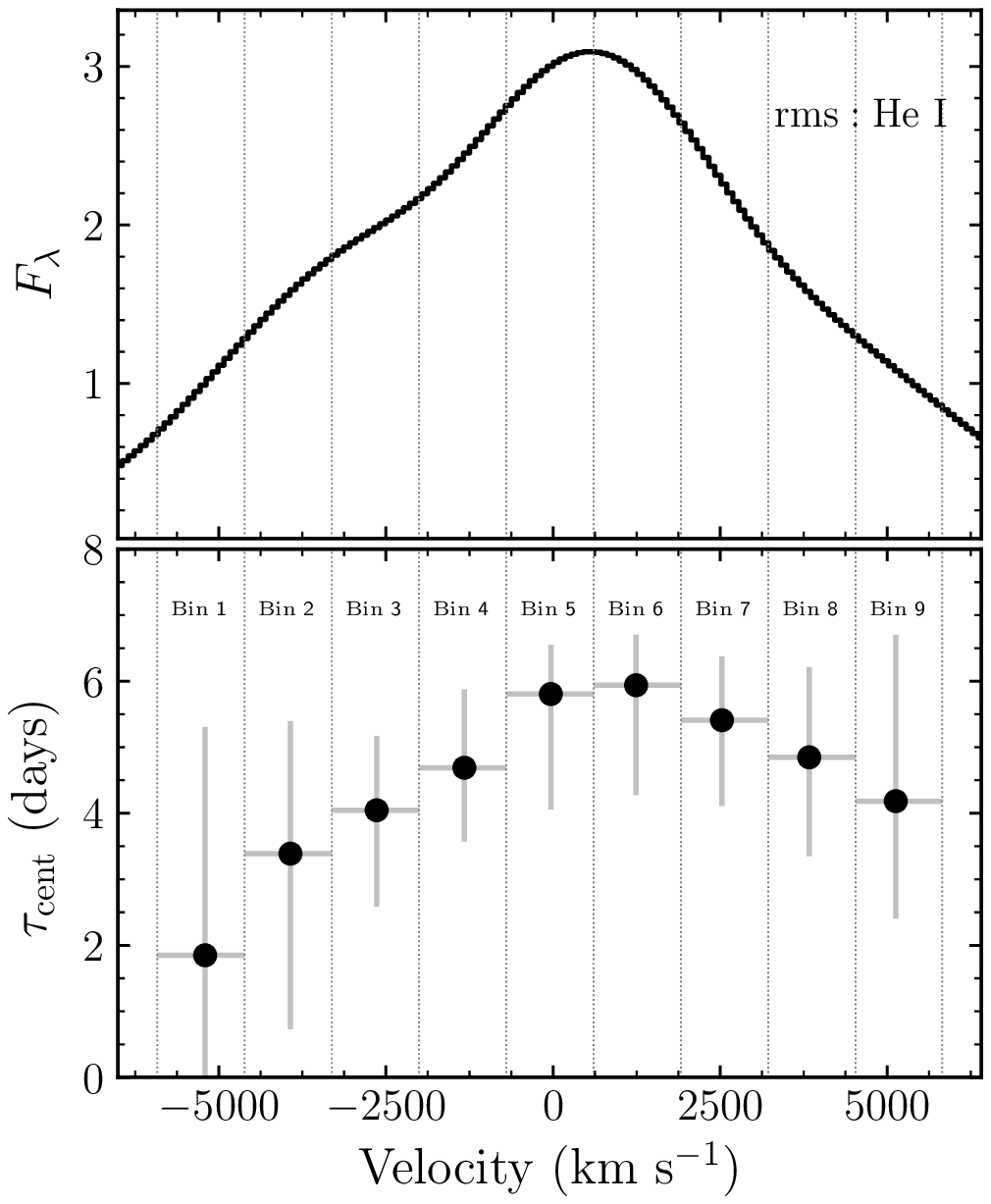}\\
\includegraphics[angle=0,width=0.44\textwidth]{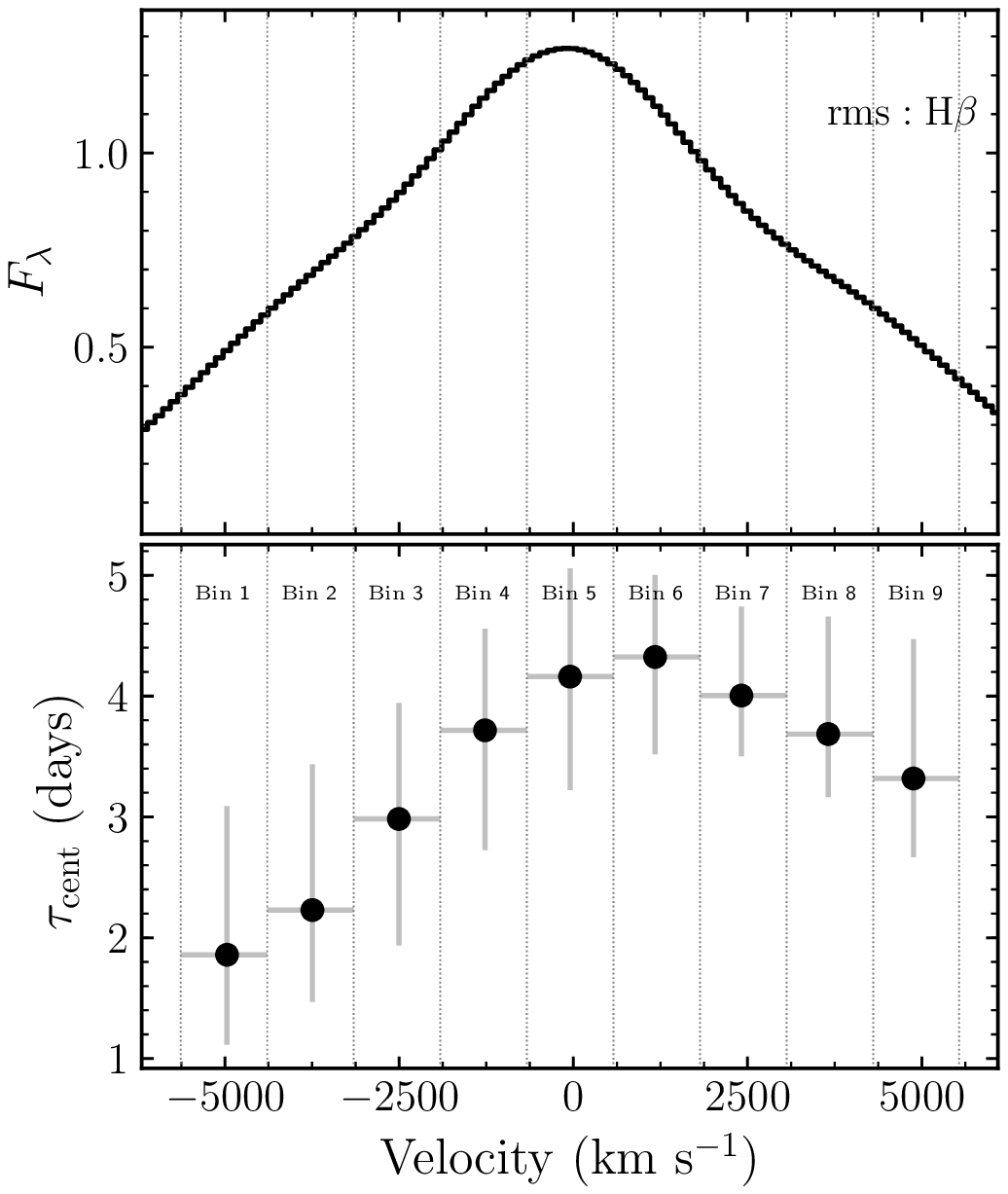}
\includegraphics[angle=0,width=0.47\textwidth]{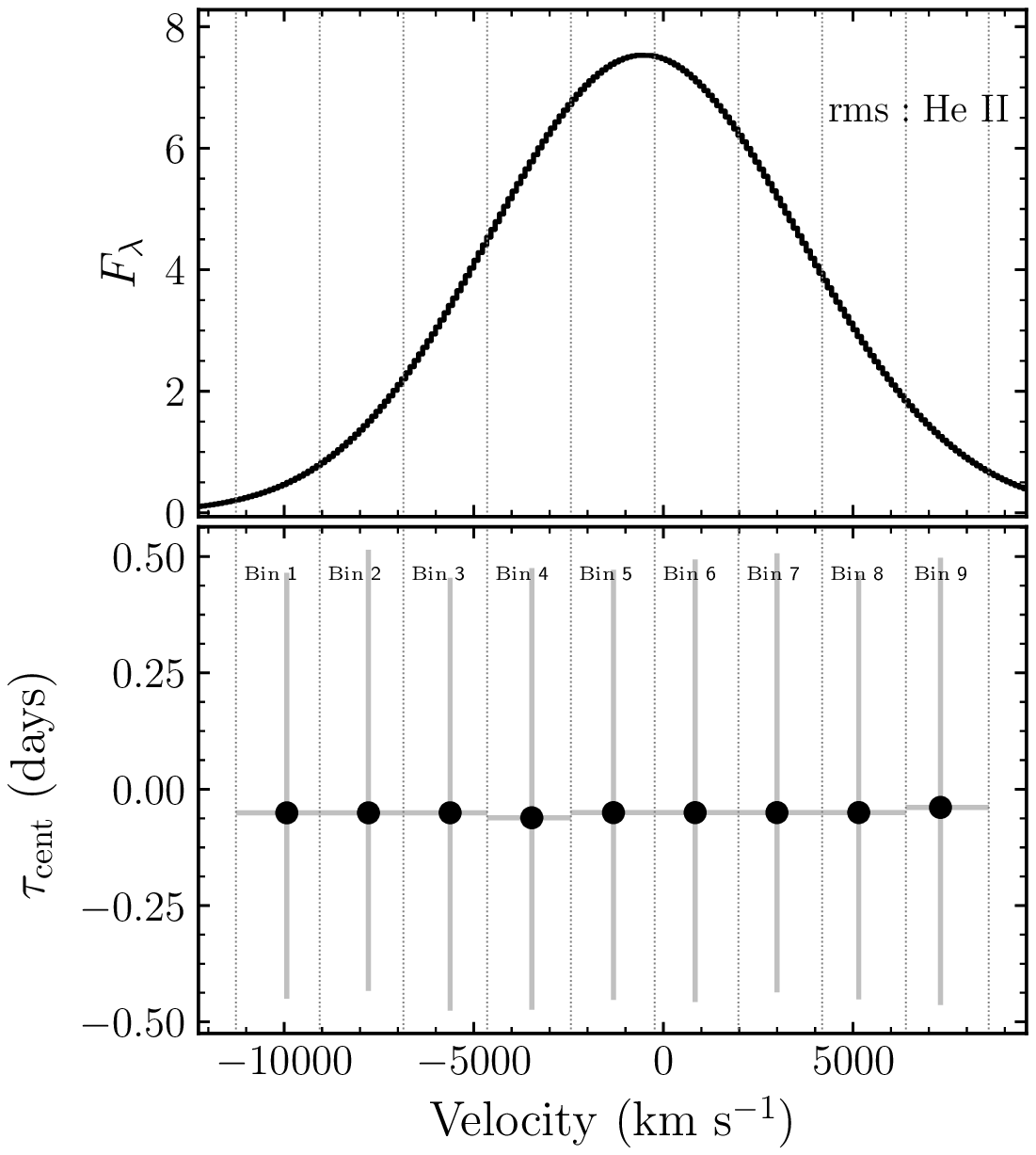}
\caption{\footnotesize
The rms profiles (top panels) and velocity-resolved lag profiles (bottom panels) of the broad H$\gamma$, 
H$\beta$, He~{\sc i}, and He~{\sc ii} lines. The vertical dash-lines are the edges of the velocity bins. 
Each bin is labeled with {\tt Bin} number (Bin 1$-$9), which is one-to-one correspondence 
with Figures~(\ref{fig_hbeta}, \ref{fig_hgamma}, \ref{fig_heii}, \ref{fig_hei}). 
}
\label{fig_vrm}
\end{figure*}

\subsection{Velocity-resolved reverberation mapping}
\label{sec_vm}
The BLR is an extended region and the velocity of the gas is most likely a function of radii. 
The BLR gas in different radii should respond to the continuum variations on slightly different delays. 
The reverberation lags measured between the continuum variations and the total fluxes of the 
broad emission lines in Table~\ref{tab_rm} represent the radii averaged by the emissivity function of the BLR. 
Based on a single broad emission line, 
velocity-resolved reverberation mapping was widely used to reveal the kinematic signatures of the BLR in many ANGs 
(e.g., \citealt{Denney2009a,Denney2010,Bentz2009b,Bentz2010,Barth2011a,Barth2011b,Grier2013,
Du2016,Lu2016,Pei2017,DeRosa2018,Zhang2019}). 
Although three RM-campaign of Mrk~79 have been performed before the present campaign, 
the study of the BLR kinematics is still blank. 
In this section, we constructed the velocity-resolved lag profiles of the broad emission lines 
(H$\beta~\lambda4861$, H$\gamma~\lambda4340$, He~{\sc ii}~$\lambda4686$, and He~{\sc i}~$\lambda5876$).   
Following previous method (e.g., \citealt{Denney2009b,Lu2016,Pei2017}), our procedure is as follows. 
At first, we calculated the rms spectrum of these broad lines, 
as illustrated in Figure~\ref{fig_vrm} (top panels). 
Then we selected a wavelength range and divided the rms profiles of broad emission lines into nine 
uniformly spaced bins (each bin has same velocity width $\sim$1300 ${\rm km~s^{-1}}$)
\footnote{The instrumental broadening ($\psi$) is 829~${\rm km~s^{-1}}$ for adopting $2.5^{\prime\prime}$ slit. 
The average broadening in each velocity bin is significantly smaller than bin width ($\sim$1300 ${\rm km~s^{-1}}$), 
which means that the instrumental broadening has a negligible impact on our analysis.}. 
The light curves of each bin were finally obtained by just integrating the fluxes in the bin, 
corresponding light curves were shown in the left panels of Figures~(\ref{fig_hbeta}, \ref{fig_hgamma}, \ref{fig_heii}, \ref{fig_hei}) 
and were numbered with {\tt Bin} number (i.e., Bin 1 to 9 from blue wing, line core to red wing of broad lines). 
Using Equation~(\ref{eq_fvar}), we calculated variability amplitudes $F_{\rm var}$ of the velocity-resolved light curves 
and noted these values in the left panels of Figures~(\ref{fig_hbeta}, \ref{fig_hgamma}, \ref{fig_heii}, \ref{fig_hei}). 
The reverberation lag of each bin and associated uncertainty were 
determined using same procedures as described in Section~\ref{sec_lag}. 
The results of cross correlation analysis are shown in the right panels of Figures~(\ref{fig_hbeta}, \ref{fig_hgamma}, \ref{fig_heii}, \ref{fig_hei}). 

Bottom panels of Figure~\ref{fig_vrm} show the velocity-resolved lag profiles (VLPs) of the broad 
H$\beta~\lambda4861$, H$\gamma~\lambda4340$, He~{\sc ii}~$\lambda4686$, and He~{\sc i}~$\lambda5876$ emission lines. 
The vertical dash-lines are the edges of the velocity bins. Each bin is labeled with {\tt Bin} number, 
which is one-to-one correspondence with Figures~(\ref{fig_hbeta}, \ref{fig_hgamma}, \ref{fig_heii}, \ref{fig_hei}). 
For the VLP of He~{\sc ii}~$\lambda4686$, the absolute value of 9 velocity-dependent delays are less than 0.5~day, 
which is shorter than median sampling of 1.0~day. 
A higher sampling is necessary to construct clear VLP of broad He~{\sc ii}~$\lambda4686$ line by decreasing the errors. 
The VLPs of H$\gamma~\lambda4340$, H$\beta~\lambda4861$, and He~{\sc i}~$\lambda5876$ 
almost have same kinematic signatures. 
They demonstrate that the high-velocity gas in the wings exhibits a shorter lag than the low-velocity gas. 
This is consistent with the viral nature of gas motion in the BLR \citep{Bentz2009b,Grier2013}, that is gas kinematics of the BLR during 
the monitoring period is dominated by Keplerian gas motion. 
However, the lag in the red wing is slightly larger than the lag in the blue wing, 
and the largest delay response occurs in the red side of the line core 
(i.e., {\tt Bin6}; $\sim$+1500~${\rm km~s^{-1}}$) for these broad emission lines. 
Similar signature has been seen in NGC~3227 (see Figure 3 of \citealt{Denney2009a}). 
This is a signature of outflow gas motion \citep{Bentz2009b,Grier2013}. 
These complicated signatures should suggest that the BLR of Keplerian motion 
in Mrk~79 probably exists the outflow gas motion during the monitoring period.

\subsection{Black hole mass and accretion rates}
Using the RM measurements of the broad emission lines, 
we determined BH mass of Mrk~79 by equation 
\begin{equation}
M_{\bullet}=f_{\rm BLR}\frac{c\tau_{_{\rm BLR}} V_{_{\rm BLR}}^2}{G} \equiv f_{\rm BLR}{\rm VP}, 
\label{eq_vp}
\end{equation}
where $c\tau_{_{\rm BLR}}$ is the size of the BLR, $c$ is the speed of light, 
$\tau_{_{\rm BLR}}$ is the lag of the broad emission line with respect to the continuum variation, 
$G$ is the gravitational constant, $V_{_{\rm BLR}}$ (i.e., FWHM or $\sigma_{\rm line}$) 
is the velocity width of the broad emission lines, and VP is commonly called the virial product. 
The coefficient $f_{\rm BLR}$ called virial factor depends on the geometry,  kinematics, and inclination of the BLR. 

We first calculated VP based on the different broad emission lines and tabulated results in Table~\ref{tab_rm}. 
\cite{Graham2011} the first evaluated the virial factor $f_{\rm BLR}$ taking into account the morphology of the host galaxies, 
and found that the factor $f_{\rm BLR}$ for barred galaxies is three times lower than that for non-barred galaxies.  
\citet{Ho2014} reevaluated the factor $f_{\rm BLR}$ for the RM AGN sample taking into account the bulge type 
of the host galaxies and found that the systematic difference in $f_{\rm BLR}$ between 
barred and non-barred galaxies qualitatively resembles the dependence on bulge type. 
\citet{Ho2014} 
suggested that pseudo-bulge notably has a lower $f_{\rm BLR}$ than in classical bulge. 
For FWHM measured from mean spectrum, 
$f_{\rm BLR}=0.5\pm0.2$ for pseudo-bulges, whereas $f_{\rm BLR}=1.3\pm0.4$ for classical bulges. 
For $\sigma_{\rm line}$ measured from rms spectrum, 
$f_{\rm BLR}=3.2\pm0.7$ for pseudo-bulges, whereas $f_{\rm BLR}=6.3\pm1.5$ for classical bulges. 
The host galaxy of Mrk~79 has a classical bulge \citep{Kim2017}. 
Based on broad H$\beta$ line, multiplying virial factor $f_{\rm BLR}=1.3\pm0.4$ and $f_{\rm BLR}=6.3\pm1.5$ to 
${\rm VP|_{FWHM_{H\beta}}}(=2.91^{+0.54}_{-0.52}\times10^{7}M_{\odot})$ and 
${\rm VP|_{\sigma_{H\beta}}}(=0.82^{+0.16}_{-0.15}\times10^{7}M_{\odot})$, 
we obtained $M_{\bullet}=3.79^{+1.36}_{-1.35}\times10^{7}M_{\odot}$ and 
$5.13^{+1.57}_{-1.55}\times10^{7}M_{\odot}$, respectively. 
Our measurement of BH mass for Mrk~79 is consistent with previous results (see \citealt{Peterson1998,Bentz2013}). 
The bulge of Mrk~79 has a stellar velocity dispersion $(130\pm12)~{\rm km~s^{-1}}$ \citep{Nelson2004}. 
Using the latest $M_{\bullet}-\sigma_{*}$ relation \cite{Kormendy2013}, 
we obtained $M_{\bullet}|{\sigma_{*}}=4.68\times10^{7}M_{\odot}$ with the scatter 0.34 dex, 
which is consistent with estimates from ${\rm VP|_{\sigma_{H\beta}}}$ and ${\rm VP|_{FWHM_{H\beta}}}$. 

Based on the standard model of accretion disk, 
the dimensionless accretion rates are related to the 5100~\AA~luminosity and BH mass via \citep{Du2015}
\begin{equation}
\mathdotM=\frac{\dot{M}_{\bullet}}{L_{\rm Edd}c^{-2}}=20.1\left(\frac{\ell_{44}}{\cos i}\right)^{3/2}M_7^{-2}, 
\label{eq_mdot}
\end{equation}
where $\dot{M}_{\bullet}$ is the mass accretion rates, $L_{\rm Edd}$ is the Eddington luminosity, $c$ is the speed of light, 
$\ell_{44}=L_{5100}/10^{44}~\ergs$ is optical luminosity at 5100~\AA, $M_7=\bhm/10^7\sunm$ is BH mass, 
$\cos i$ is the cosine of the inclination of the accretion disk. 
$i=24$ degrees for Mrk~79 \citep{Gallo2011}. 
Using $M_{\bullet}=5.13^{+1.57}_{-1.55}\times10^{7}M_{\odot}$ and 
$L_{5100}=(1.45\pm0.14)\times10^{43}~{\rm erg~s^{-1}}$, 
we obtained accretion rates $\dot{M}_{\bullet}=(0.05\pm0.02)~L_{\rm Edd}~c^{-2}$, 
indicating that Mrk~79 is a sub-Eddington accreting AGN. 

\begin{figure*}[ht!]
\centering
\includegraphics[angle=0,width=0.95\textwidth]{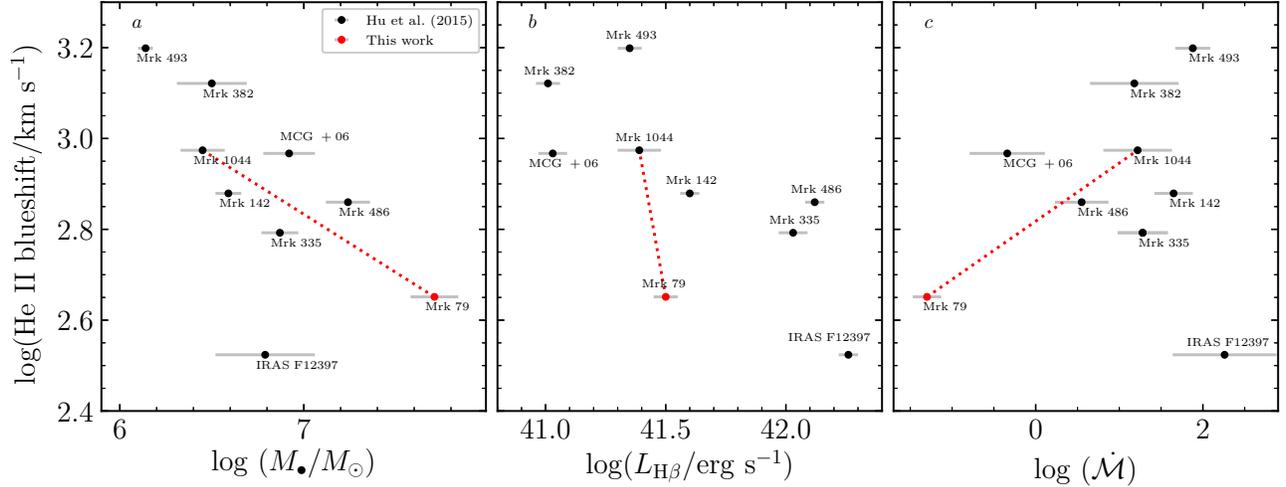}
\caption{\footnotesize
Relation between blueshift velocity of the broad He~{\sc ii}~$\lambda4686$ line and AGN properties 
including BH mass, luminosity of the broad H$\beta$ line, and accretion rates. 
Disk wind was just detected in Mrk~79 and Mrk~1044, they are connected in a dashed line. 
}
\label{fig_heshift}
\end{figure*}

\section{Discussion}
\label{sec_dis}
\subsection{Indirect evidence of the BLR as a disk wind} 
On the one hand, 
UFOs and warm absorbers are identified in term of X-ray spectrum for Mrk~79 \citep{Tombesi2010,Tombesi2011,Gallo2011}, 
but their geometries remain unclear \citep{Parker2018}, 
We don't know whether blueshifted UV absorbers/emitters exist in Mrk~79 in absence UV spectrum. 
In Section~\ref{sec_meanrms}, we detected marginal blueshift of the broad He~{\sc ii}~$\lambda4686$ line 
with velocity of $\sim \rm 450~km~s^{-1}$, which indicates an outflow of high-ionization gas (e.g., He~{\sc ii} emitters). 
But it is significantly slower than the velocity of UFOs ($0.016~V_{\rm UFOs,out}$). 
As suggestion of disk wind model \citep{Murray1995,OBrien2005,Tombesi2013}, 
this result may attribute to a possibility that the X-ray outflow could be the source of some of the BLR gas. 
In addition, based on the velocity-resolved lag profiles of broad 
He~{\sc i}~$\lambda 5876$, H$\beta~\lambda4861$, and H$\gamma~\lambda4340$ lines (Section~\ref{sec_vm}),  
we found that low-ionization gas of the BLR exhibits the outflowing signature during the monitoring period. 
Meanwhile, we found that Mrk~79 is similar to NGC~3227 (see Section~\ref{sec_intro}) in some aspects. 
For example, (1) UFOs and warm absorbers were detected in both AGNs; 
(2) the low-ionization BLR of both AGNs exhibit the outflowing component (see Section~\ref{sec_vm} and \citealt{Denney2010}). 
While, based on velocity-resolved RM, 
we do not find that the BLR doubtlessly exhibits an outflowing component for normal AGN (i.e., no disk wind) so far. 
These findings may indicate that the outflowing BLR could be associated with disk wind, 
and may support the notion that disk wind could be the source of some of the BLR gas as suggested by \cite{OBrien2005}. 
All of these phenomenons including UFOs, warm absorbers, and the kinematics of the high- and low-ionization BLR, 
may provide an indirect evidence that the BLR of Mrk~79 probably originates from disk wind. 
However, simultaneous observations of multi-band spectra are necessary to further constrain this speculation. 

It should be noted that, NGC~3227 as one of candidates of UFOs, 
the outflowing BLR in 2007 \citep{Denney2010} turns to virialized BLR in 2012. 
\cite{DeRosa2018} suggested that the most likely reason for this change is that 
the BLR structure is probably complex and consists of multiple components---a disk and a wind. 
In this case, the decelerating and cooling outflow may gradually fall and turn to virialization, 
or it could be real intrinsically that the BLR kinematics are variable on the dynamical timescale 
as we saw in NGC~5548 \citep{Lu2016,Kriss2019ApJ,DeRosa2018,Xiao2018}. 
Therefore, the different BLR kinematics will be observed from the different campaigns. 

On the other hand, when NGC~5548 is seeming to come out of faint state (e.g., \citealt{DeRosa2018,Pei2017,Kriss2019ApJ}), 
the broad emission lines failed to respond to variations in the continuum flux as the BLR `holiday'. 
\cite{Dehghanian2019} argued that X-ray absorption (observed by \citealt{Mehdipour2016}), 
produced transient obscurer, was present throughout the BLR `holiday' of NGC~5548. 
Based on the X-ray and UV band monitoring of NGC~5548, 
\cite{Kriss2019ApJ} combined the observational facts including 
the obscurer and the departure of NGC 5548's BLR from the radius$-$luminosity relationship \citep{Peterson2002,Pei2017,DeRosa2018}, 
and suggested that the obscurer is a manifestation of a disk wind launched in the brightening state. 
Coincidentally, 
what has been happening in better-studied NGC~5548 seems to has been happening in Mrk~79 as well. 
Both are seeming to come out of faint states, the H$\beta$ lags are too short for the low luminosity state, 
and both are suspected of triggering disk wind. 
Unfortunately, we don't have UV or X-ray spectra of Mrk 79. 

It is worth noting that outflow of RQ AGN should be jointly triggered and controlled by central gravity and magnetic or (and) radiation pressure. 
In this case, the velocity of outflow should anticorrelate with BH mass, 
and positively correlate with magnetic or (and) radiation pressure. 
We compiled blueshifted velocity of the broad He~{\sc ii}~$\lambda4686$ emission line from \cite{Hu2015} and this work, 
and investigated relationship between the blueshifted velocity of He~{\sc ii}~$\lambda4686$ emitters and AGN properties including BH mass, 
luminosity of broad H$\beta$ line (which is used as proxy of UV luminosity), and accretion rates in Figure~\ref{fig_heshift}. 
It should be noted that, in this small sample, disk wind was only detected in Mrk~79 and Mrk~1044 \citep{Parker2018}. 
This sample shows a possible trend that the blueshifted velocity of He~{\sc ii} emitters are anticorrelated with BH mass, 
which may suggest that central gravity plays a potential role on the terminal velocity of outflow. 
While we cannot see clear trend between blueshifted velocity of He~{\sc ii} emitters 
and the rest of AGN properties (Figure~\ref{fig_heshift}{\it b}~and~\ref{fig_heshift}{\it c}). 
An larger sample is necessary to responsibly investigate these relations. 
On the other hand, based on the results of 18 RM-campaigns of NGC~5548, \cite{Lu2016} 
recently found that the BLR size ($\tau_{\rm BLR}$) 
follows the variation of optical luminosity in the long-term timescale for NGC~5548 but 
with a time delay $\rm {\sim 3~yr}$ (also see \citealt{Kriss2019ApJ}), and implied the potential role of radiation pressure. 
Mrk~79 is monitored by 4 RM campaigns so far (including this work), nevertheless, 
more and dense RM campaigns are necessary to investigate this nature. 

In addition, many possible origins of the BLR are developed besides the above case. 
For example, a series of works suggested that part of the BLR gas (or broad emission line) originate from the outer region of accretion disk 
(e.g., \citealt{Collin-Souffrin1987,Collin-Souffrin1990}), 
but this scenario only produces low-ionization emission lines; 
\cite{Wang2017} modelled H$\beta$ profile using dynamical model of different type clouds and suggested that 
tidally disrupted clumps from the torus may represent the source of the BLR; 
\cite{Czerny2011} suggested that the BLR is a failed dusty wind from the outer accretion disk (also see \citealt{Czerny2017}); 
\cite{Baskin2018} suggested the BLR originate from the dusty inflated accretion disk (see Figure 13 of \citealt{Baskin2018}). 
However, whether these possible origins of the BLR are an intermediate state (phase) of multiphase disk wind 
should be studied in the future.

\begin{figure}[ht!]
\includegraphics[angle=0,width=0.5\textwidth]{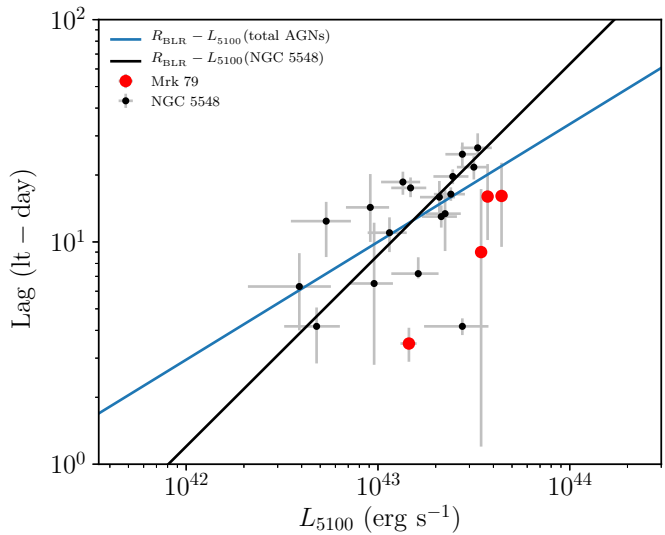}
\caption{\footnotesize
The $R_{\rm BLR}-L_{\rm 5100}$ relation of Mrk~79 (red circles) along with NGC~5548 (black circles). 
Canonical $R_{\rm BLR}-L_{\rm 5100}$ relationship (blue line) is taken from \cite{Bentz2013}. 
The $R_{\rm BLR}-L_{\rm 5100}$ relationship (black line) of NGC~5548 is taken from \cite{Lu2016}. 
}
\label{fig_rl}
\end{figure}

\subsection{Radius$-$Luminosity relation}
\label{sec_rl}
Many works focused on investigating the kinematics of the BLR in NGC~5548, and constructed its 
$R_{\rm BLR}-L_{\rm 5100}$ relationship \citep{Peterson2004,Bentz2007,Denney2010,Lu2016,Pei2017,DeRosa2018}. 
Comparing the observation features of Mrk~79 in the past RM campaigns (see Section~\ref{sec_result}) with NGC~5548, 
we seemingly saw another `NGC~5548-like' AGN in some aspects. For example, 
(1) the velocity widths of broad emission lines  (such as H$\beta$ line) are very broad (${\rm >4000~km~s^{-1}}$), 
that is they are broad-line Seyfert galaxy; 
(2) They have a classical bulge in the host galaxy \citep{Kim2017}; 
(3) Their BLR size and optical luminosity exhibit large variation (see Table~\ref{tab_opl}; \citealt{Lu2016,Kriss2019ApJ}). 
Mrk~79 is expected to be a new candidate can be used to construct another $R_{\rm BLR}-L_{\rm 5100}$ 
relationship of individual object because of enormous change of AGN continuum  
and the BLR size (Table~\ref{tab_opl}). 
Figure~\ref{fig_rl} presents the canonical $R_{\rm BLR}-L_{\rm 5100}$ relationship (slope$=$0.53) from \cite{Bentz2013} 
along with $R_{\rm BLR}-L_{\rm 5100}$ relationship of NGC~5548 (slope$=$0.86, \citealt{Lu2016}). 
Black dots with error bars display the $R_{\rm BLR}-L_{\rm 5100}$ relation of NGC~5548 reverberation, 
red circles with error bars display the $R_{\rm BLR}-L_{\rm 5100}$ relation of Mrk~79 reverberation. 
We found that the current $R_{\rm BLR}-L_{\rm 5100}$ relation of Mrk~79 deviates from the  
canonical $R_{\rm BLR}-L_{\rm 5100}$ relationship (blue line) 
and NGC~5548's $R_{\rm BLR}-L_{\rm 5100}$ relationship (black line). 

We noted that the previous works have described how the BLR radius in an individual object 
changes as the mean optical luminosity varies with time 
(e.g., \citealt{Baldwin1995,Peterson1999,Peterson2002}). Based on the NGC5548's results of 13-season RM campaigns, 
\cite{Peterson2002} found that the observed relation between the BLR radius and the luminosity 
in an individual AGN (Figure 5 of \citealt{Peterson1999} and Figure 3 of \citealt{Peterson2002}) 
is consistent with the prediction of the simple photoionization equilibrium model. 
However, the latest studies found that (1) the BLR radius correlates with the averaged luminosity of AGN but with a delay, 
and this delay coincides with the BLR's dynamical timescale \citep{Ulrich1991,Lu2016,Kriss2019ApJ}; 
(2) The shortened H$\beta$ lags correlate with the accretion rates of AGN (e.g., \citealt{Du2018}). 
These results may imply that the BLR physics is probably complicated, 
the evolution of the BLR structure and kinematics may change the correlation 
between the BLR radius and luminosity of AGN predicted by photoionization equilibrium model. 
It is possible that comparing the $R_{\rm BLR}-L_{\rm 5100}$ relationships of different AGNs 
(e.g., NGC~5548, Mrk~79 and so on) in the near future 
and investigating their similarities and differences 
could help us to understand the scatter ($\sim~0.3$~dex, see \citealt{Bentz2013,Du2018,Grier2017}) 
of canonical $R_{\rm BLR}-L_{5100}$ relationship. 

\section{Summary}
\label{sec_con}
We developed an monitoring project to investigate 
the kinematics of the BLR in AGN with UFOs 
and explore potential connection between the BLR and disk wind. 
This is the first result from a new RM-campaign of Mrk~79, 
which was undertaken by Lijiang 2.4-m telescope (LJT+YFOSC). 
Spectral fitting scheme was adopted to remove the host-galaxy component from spectrum 
and improve the measurement quality of light curves. 
Reverberation analysis of several broad emission lines 
(H$\beta~\lambda4861$, H$\gamma~\lambda4340$, He~{\sc ii}~$\lambda4686$, and He~{\sc i}~$\lambda5876$) 
are carried out. Based on the present campaign, We obtained the following results. 

\begin{enumerate}
\item 
Mrk~79 is seeming to come out the faint state, 
the average flux of AGN at 5100~\AA~approximates a magnitude fainter than previous record holder. 
We found that the variability amplitudes of the broad emission lines meet 
$F_{\rm var, He~{\sc II}}  > F_{\rm var, H\gamma}  > F_{\rm var, He~{\sc I}} \ge F_{\rm var, H\beta}$ relation. 
\item
High-ionization line of He~{\sc ii}~$\lambda4686$ 
has a marginal blueshift with velocity $\sim \rm 450~km~s^{-1}$, 
which indicates an outflow of high-ionization gas. 
\item 
We successfully measured the time delays of the broad 
H$\beta~\lambda4861$, H$\gamma~\lambda4340$, He~{\sc ii}~$\lambda4686$, and He~{\sc i}~$\lambda5876$ emission lines 
with respect to the continuum variation. 
The optimum lags of H$\gamma~\lambda4340$, H$\beta~\lambda4861$, and He~{\sc i}~$\lambda5876$ lines  
marginally show a ionization stratification of the BLR. 
The lag of He~{\sc ii}~$\lambda4686$ line approximates zero, which is consistent with previous results. 
\item 
We simultaneously obtained the velocity-resolved lag profiles of the broad 
H$\gamma~\lambda4340$, H$\beta~\lambda4861$, and He~{\sc i}~$\lambda5876$ emission lines for the first time, 
which almost show same kinematic signatures. 
Specifically, 
the high-velocity gas in the wings exhibits a shorter lag than the low-velocity gas. 
However, the lag in the red wing is slightly larger than the lag in the blue wing, 
and the largest lag occurs in the red side. These complicated signatures should suggest 
that the BLR of Keplerian motion in Mrk 79 probably exists the outflow gas motion during the monitoring period.
\item 
Based on the velocity width and time delay of broad H$\beta~\lambda4861$ line, 
we measured BH mass of $M_{\bullet}=5.13^{+1.57}_{-1.55}\times10^{7}M_{\odot}$ for Mrk~79. 
This value is consistent with the estimate of $M_{\bullet}-\sigma_{*}$ relation. 
Using this BH mass and optical luminosity at 5100~\AA~$L_{5100}=(1.45\pm0.14)\times10^{43}~{\rm erg~s^{-1}}$, 
we estimated accretion rates of $\dot{M}_{\bullet}=(0.05\pm0.02)~L_{\rm Edd}~c^{-2}$. 
Mrk 79 is a sub-Eddington accreting AGN. 
\item 
We found that the current $R_{\rm BLR}-L_{\rm 5100}$ relation of Mrk~79 reverberation deviates 
from the canonical $R_{\rm BLR}-L_{\rm 5100}$ (slope$=$0.53) and NGC~5548's $R_{\rm BLR}-L_{\rm 5100}$ (slope$=$0.86) relationships. 
More and dense RM campaigns are necessary to construct robust $R_{\rm BLR}-L_{\rm 5100}$ relationship of Mrk~79. 
\end{enumerate}

As discussed in Section~\ref{sec_dis}, 
although we don't know whether blueshifted UV absorbers/emitters exist in Mrk~79 in absence UV spectrum, 
many findings including UFOs, warm absorbers , and the BLR kinematics of the high- and low-ionization gas, 
indicate that the BLR of Mrk~79 probably originates from disk wind launched from accretion disk. 
Nevertheless, simultaneous observations of multi-band spectra are necessary to confirm this speculation. 

\acknowledgements
{We are grateful to the referee for useful suggestions that improved the manuscript. 
We acknowledge the support of the staff of the Lijiang 2.4-m telescope. 
Funding for the telescope has been provided by CAS and the People's Government of Yunnan Province. 
This research is supported in part by National Key Program for Science and Technology Research and Development of China
(grants 2016YFA0400701 and 2016YFA0400702). 
L.C.H. acknowledges financial support from the National Natural Science Foundation of China (NSFC; 11721303).} 
K.X.L. acknowledges financial support from the NSFC (11703077) and from the Light of West China Program provided by CAS (No. Y7XB016001). 
MK was supported by the National Research Foundation of Korea (NRF) grant funded by the Korea government (MSIT) (No. 2017R1C1B2002879). 
W.H.B. acknowledges financial support from the NSFC (11973029).


\appendix
\section{Insight into the Spectroscopy and Calibration} 
\label{sec_a}
\begin{figure*}[ht!]
\centering
\includegraphics[angle=0,width=0.85\textwidth]{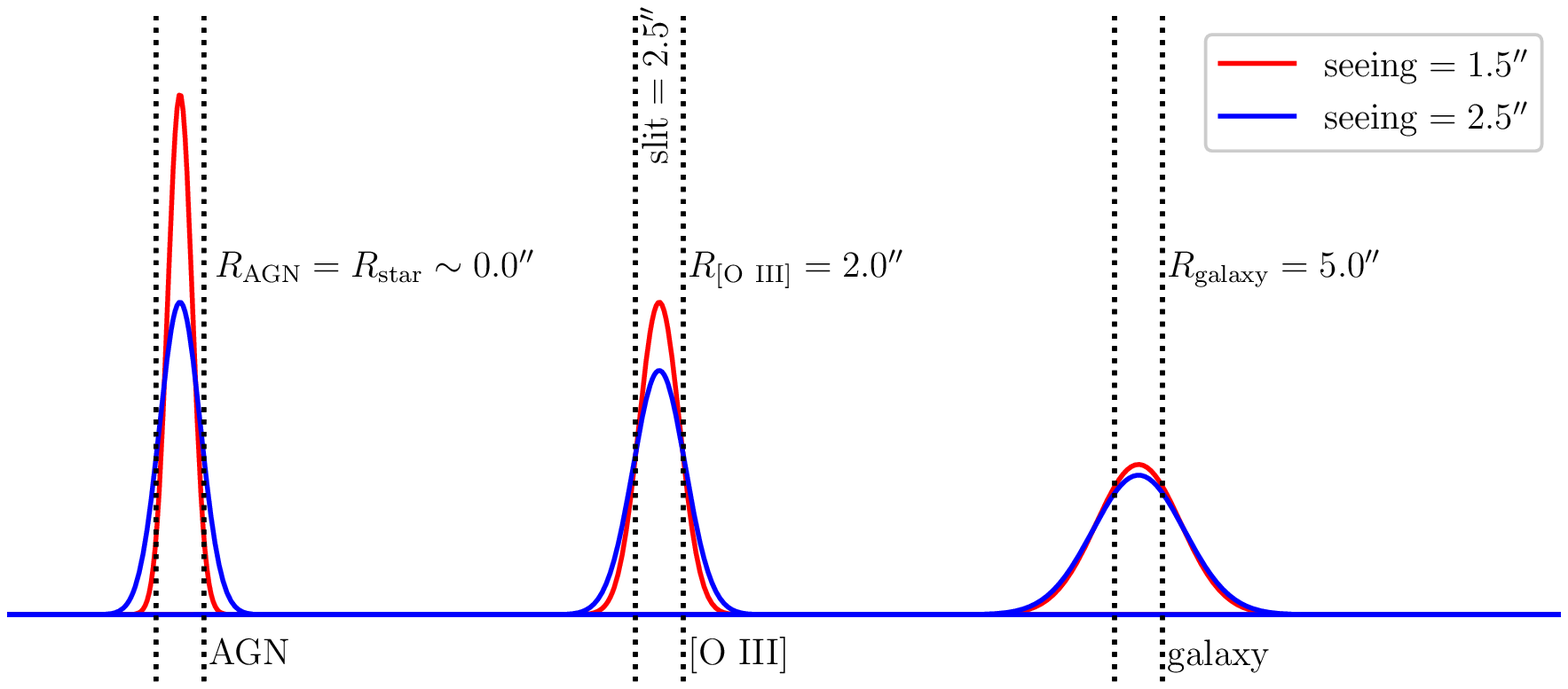}
\caption{\footnotesize 
Schematic diagram for showing the flux variation in the slit. 
Each pair of dotted lines represent the slit (width$=2.5^{\prime\prime}$) adopted in spectroscopy. 
The Gaussian profiles (left to right) represent the flux distribution (i.e., surface-brightness distribution) of the point sources (AGN, the BLR and star), 
the slightly extended source ([O~{\sc III}]) and the very extended source (host galaxy), respectively. 
The red profiles are broadened by seeing of $1.5^{\prime\prime}$, 
the red profiles are broadened by seeing of $2.5^{\prime\prime}$. 
$R_{\rm star}$, $R_{\rm AGN}$, $R_{\rm [O~III]}$ and $R_{\rm galaxy}$ roughly represent the radii of 
the comparison star,  AGN, [O~{\sc iii}] emission region (refer to \citealt{Schmitt2003a}) and the host galaxy (refer to \citealt{Bentz2009a}), respectively. 
}
\label{fig_sch}
\end{figure*}

\begin{figure}[ht!]
\includegraphics[angle=0,width=0.5\textwidth]{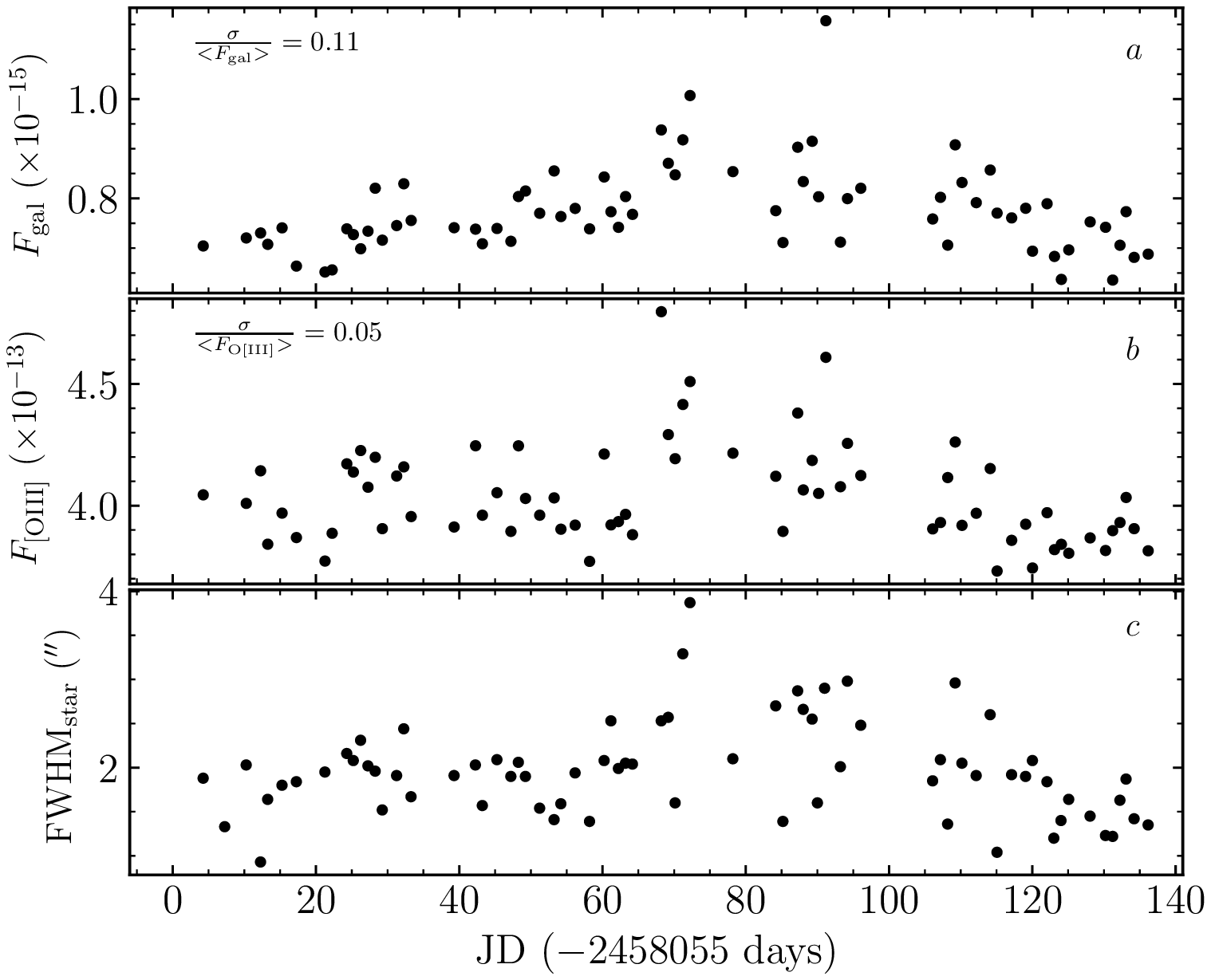}
\caption{\footnotesize 
Panels ({\it a-b}) are the apparent variation in fluxes of the host galaxy and [O~{\sc iii}]~$\lambda5007$ 
measured from the best-fit components. 
The width (${\rm FWHM_{star}}$) of star's flux distribution was measured from the short exposure image, 
the image was observed before and near the spectroscopy. 
Panel ($c$) shows the variation of ${\rm FWHM_{star}}$. 
}
\label{fig_elc}
\end{figure}

\begin{figure}[ht!]
\includegraphics[angle=0,width=0.45\textwidth]{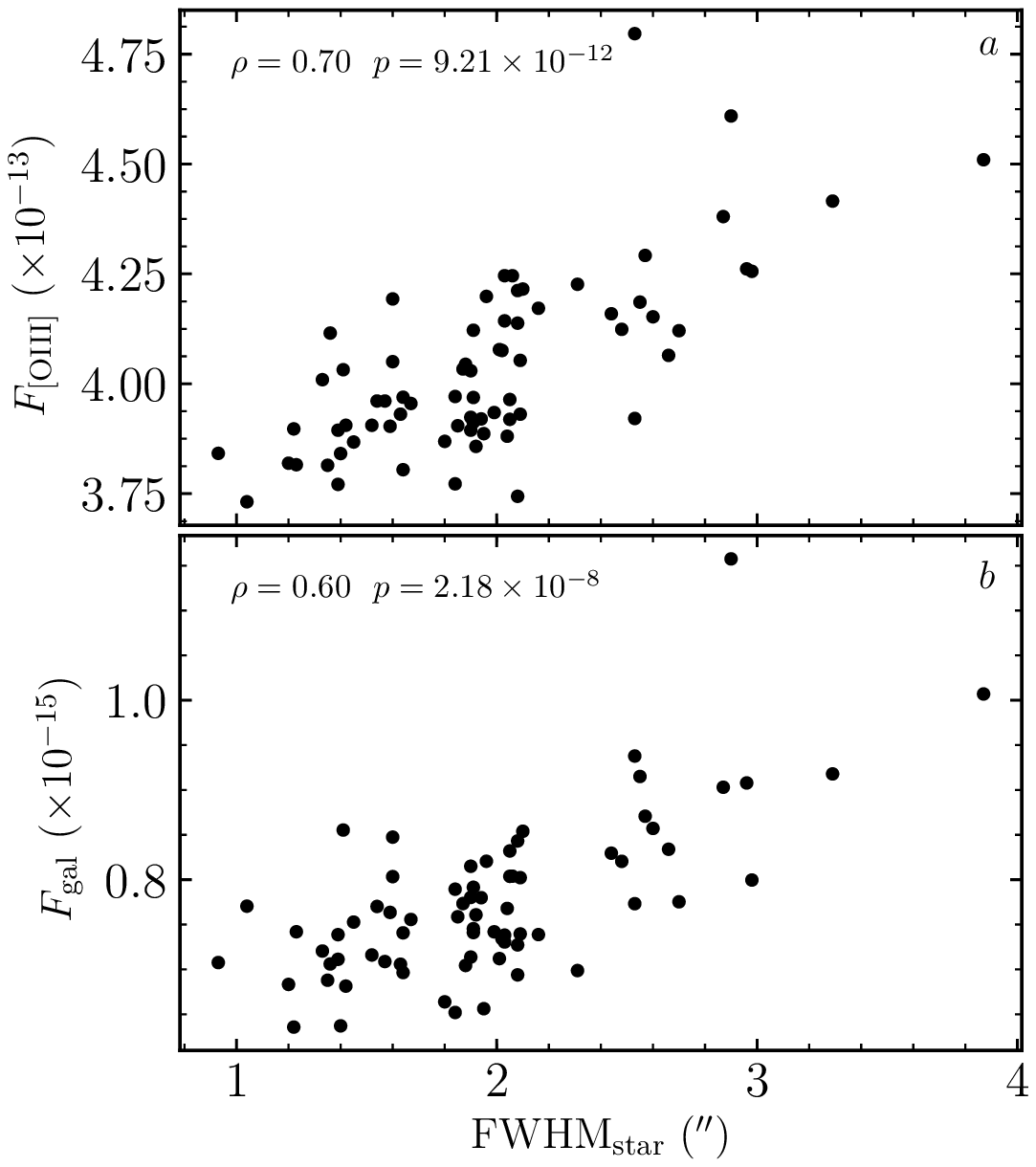}
\caption{\footnotesize 
Relation between the fluxes of extended components ($F_{\rm [O~{\sc III}]}$ and $F_{\rm gal}$) and ${\rm FWHM_{star}}$ variation. 
Spearman rank-order correlation coefficient ($\rho$) and the p-value ($p$) are noted in the panels. 
}
\label{fig_seecor}
\end{figure}

To qualitatively study the apparent variation in flux of the host galaxy and the [O~{\sc iii}]~$\lambda5007$ 
mentioned in Section~\ref{sec_meanrms}~and~\ref{sec_lc}, 
we have insight into the details of spectroscopy and flux calibration in this Appendix. 
Mrk~79 consist of the AGN (including the BLR), the host galaxy and the [O~{\sc iii}] emission region (i.e., NLR: the narrow-line region), 
which is observed along with the comparison star simultaneously. 
At first, we simply considered the size of different components. 
The AGN including the BLR along with the comparison star are point sources, 
the intrinsic size of point source approximates $0^{\prime\prime}$ in remote distance. 
For the NLR and the host galaxy, we noted that \cite{Peterson1995} have produced 
the models of the surface-brightness distribution of the NLR and 
host-galaxy distribution from ground-based images in NGC 4151 and NGC 5548, and found that 
(1) the NLR of NGC 5548 is point-like source; 
(2) The NLR of NGC 4151 is slightly extended source. 
Based on the growth curve for the [O~{\sc iii}]~$\lambda5007$ flux distribution constructed by \cite{Peterson1995}, 
we estimated the radius (defined by 80\% of the integrated [O~{\sc iii}]~$\lambda5007$ flux) 
of the [O~{\sc iii}] emission region in NGC 4151 ($\sim 1.5^{\prime\prime}$); 
(3) The host galaxy is very extended source, 
this is consistent with the results of \textit{HST} image decomposition for the local AGNs (see \citealt{Bentz2009a,Kim2017}). 
In Mrk~79, based on \textit{HST} image of the [O~{\sc iii}], 
\cite{Schmitt2003a} measured the effective radius (defined by 50\% of the integrated [O~{\sc iii}]~$\lambda5007$ flux), 
extent of photometric semi-major and semi-minor axes of the [O~{\sc iii}] emission (see Table 3 of \citealt{Schmitt2003a}). 
The semi-major axes, which is roughly perpendicular to our long slit (see Figure 8 of \citealt{Schmitt2003a}, 
and the position angle of our long slit is $-99^{\circ}$), has a size of $2.1^{\prime\prime}$, comparable to the width of slit. 
This valuable measurement shows the NLR of Mrk~79 along with a part of other AGNs is slightly extended source. 
For AGNs with a slightly extended NLR observed with a broad spectrograph slit of $5.0^{\prime\prime}$ 
(adopted by many previous RM campaigns, see Table 12 of \citealt{Bentz2013}), standard spectral calibration method using the 
[O~{\sc iii}]~$\lambda5007$ as calibrator provides precise internal flux calibration of spectra (see \citealt{Fausnaugh2017,Peterson1995}), 
because broad spectrograph slit integrally observed [O~{\sc iii}]~$\lambda5007$ emission of AGNs. 
This is not the case for a small spectrograph slit ($2.5^{\prime\prime}$), 
the seeing-induced aperture effects will cause the apparent variation in flux of extended source. 
We give a quantitative analysis about this point in the next, in fact, 
which is similar to \cite{Peterson1995}'s analysis of aperture effects on the accuracy of ground-based spectrophotometry.

Figure~\ref{fig_sch} is a spectroscopic schematic diagram for showing the flux variation of different components in the slit. 
We use red and blue Gaussian profiles to present the surface-brightness distribution of these components broadened by different seeing 
(in practice, the slightly or very extended source has a flatter surface-brightness distribution than Gaussian profile, see \citealt{Peterson1995}). 
For each component, the area surrounded by red and blue Gaussians within the small slit present 
the fractions of light loss caused by varying seeing. 
Figure~\ref{fig_sch} shows that 
the fractions of light loss due to varying observing conditions (e.g., seeing) are dependent on the size 
of the object (i.e., the width of surface-brightness distribution), seeing and the width of slit (also see \citealt{Peterson1995}). 
If we use $\psi^{\rm star}$, $\psi^{\rm AGN}$, $\psi^{\rm [O~III]}$, and $\psi^{\rm gal}$ to represent the 
percentage of flux remaining in the aperture for different components, 
the observed flux can be described by 
\begin{equation}
\begin{split}
\label{eqn1}
F_{\rm obs}^{\rm star} = \psi^{\rm star} \times F_{\rm abs}^{\rm star} 
\end{split}
\end{equation}
for the comparison star, 
\begin{equation}
\begin{split}
\label{eqn1}
F_{\rm obs}^{\rm AGN} = \psi^{\rm AGN} \times F_{\rm abs}^{\rm AGN}
\end{split}
\end{equation}
for the AGN, 
\begin{equation}
\begin{split}
\label{eqn1}
F_{\rm obs}^{\rm [O~III]} = \psi^{\rm [O~III]} \times F_{\rm abs}^{\rm [O~III]}
\end{split}
\end{equation}
for [O~{\sc iii}]~$\lambda5007$ (from the extended NLR), and 
\begin{equation}
\begin{split}
\label{eqn1}
F_{\rm obs}^{\rm gal} = \psi^{\rm gal} \times F_{\rm abs}^{\rm gal}
\end{split}
\end{equation}
for the host galaxy, where $F_{\rm abs}$ are the absolute fluxes of these components.  

The observed flux $F_{\rm obs}$ is corrected by multiplying to the flux-calibration factor, 
the factor is obtained by comparing the absolute flux to the observed flux of the comparison star 
(\citealt{Maoz1990, Kaspi2000, Du2014}), that is the calibrated flux 
\begin{equation}
\begin{split}
\label{eqn1}
F_{\rm cal} = F_{\rm obs} \times \frac {F_{\rm abs}^{\rm star}} {F_{\rm obs}^{\rm star}}=\frac {1}{\psi^{\rm star}} \times F_{\rm obs}. 
\end{split}
\end{equation}
Therefore, we deduced the calibrated flux of the AGN 
\begin{equation}
\begin{split}
\label{eqn1}
F_{\rm cal}^{\rm AGN} &=\frac {1}{\psi^{\rm star}} \times F_{\rm obs}^{\rm AGN}\\
&= \frac {\psi^{\rm AGN}} {\psi^{\rm star}} \times F_{\rm abs}^{\rm AGN}=f_{\rm cal}^{\rm AGN} \times F_{\rm abs}^{\rm AGN}, 
\end{split}
\end{equation}
the extended NLR 
\begin{equation}
\begin{split}
\label{e7}
F_{\rm cal}^{\rm [O~III]} &=\frac {1}{\psi^{\rm star}} \times F_{\rm obs}^{\rm [O~III]} \\
 &= \frac {\psi^{\rm [O~III]}} {\psi^{\rm star}} \times F_{\rm abs}^{\rm [O~III]}=f_{\rm cal}^{\rm [O~III]} \times F_{\rm abs}^{\rm [O~III]}, 
\end{split}
\end{equation}
and the host galaxy 
\begin{equation}
\begin{split}
\label{e8}
F_{\rm cal}^{\rm gal} &=\frac {1}{\psi^{\rm star}} \times F_{\rm obs}^{\rm gal} \\
&= \frac {\psi^{\rm gal}} {\psi^{\rm star}} \times F_{\rm abs}^{\rm gal}=f_{\rm cal}^{\rm gal} \times F_{\rm abs}^{\rm gal}.
\end{split}
\end{equation}
Where $f_{\rm cal}$ are the flux-calibration factors of different components. 

Figure~\ref{fig_sch} also shows that two point sources (the comparison star and AGN) kept in a line parallel to the slit, 
the fractions of light loss due to varying seeing are identical. 
However, the extended component (the [O~[{\sc iii}] emission region and the host galaxy) in the same slit, 
the fractions of light loss due to varying seeing are less than point source 
(similar analysis about aperture effects are addressed in Section 2.1 of  \citealt{Peterson1995}). 
In this case, for Mrk 79, the radii of different components including the AGN, the NLR, and the host galaxy meet 
$R_{\rm AGN}(\sim 0^{\prime\prime})<R_{\rm NLR}(\sim 2^{\prime\prime})<R_{\rm gal}(\sim 5^{\prime\prime})$ relation.  
When seeing increases, the percentages of flux remaining in the aperture meet 
$\psi^{\rm AGN}=\psi^{\rm star}<\psi^{\rm [O{\rm III]}}<\psi^{\rm gal}$ relation. 
The result is that the flux-calibration factors meet 
$1=f_{\rm cal}^{\rm AGN}<f_{\rm cal}^{\rm [O~III]}<f_{\rm cal}^{\rm gal}$ relation. 
That is the most extended component has the largest flux calibration factor. 
Consequently, 
(1) the flux-calibration factor of extended component should correlate with seeing, 
that is the calibrated fluxes including $F_{\rm cal}^{\rm [O~III]}$ and 
$F_{\rm cal}^{\rm gal}$ (Equation~\ref{e7} and~\ref{e8}) should correlate with seeing 
because the absolute flux (including $F_{\rm abs}^{\rm [O~III]}$ and 
$F_{\rm abs}^{\rm gal}$) is constant;  
(2) the host-galaxy fluxes ($F_{\rm cal}^{\rm gal}$) 
should be more scatter than [O~{\sc iii}]'s fluxes ($F_{\rm cal}^{\rm [O~III]}$) 
since $f_{\rm cal}^{\rm gal}$ is larger than $f_{\rm cal}^{\rm [O~III]}$ with varying seeing. 

In order to test the above analysis results from the perspective of observation, 
we measured the width (${\rm FWHM_{star}}$) of the star's flux distribution from the short exposure 
image observed before and near the spectroscopy, 
and showed the variation of ${\rm FWHM_{star}}$ in Figure~\ref{fig_elc}~(panel~$c$). 
In practice, seeing could change from one exposure to the next, 
${\rm FWHM_{star}}$ is mainly modulated by varying seeing. 
Figure~\ref{fig_elc} shows that 
the apparent variations in fluxes of the extended components (including [O~{\sc iii}] emission region and the host galaxy) 
are similar to the variation of ${\rm FWHM_{star}}$, 
and the host-galaxy fluxes (11\%) are more scatter than [O~{\sc iii}]'s fluxes (5\%). 
Figure~\ref{fig_seecor} clearly shows that the fluxes 
of extended components correlate with the variation of ${\rm FWHM_{star}}$. 
Actually, these examination results are consistent with above analysis, 
and show that the [O~{\sc iii}] emission region in Mrk~79 is slightly extended source. 
Therefore, for a narrow slit,  
the varying observing conditions will give rise to the apparent variation in flux of the extended components. 
So that the [O~{\sc iii}] remains in the rms spectrum and its fluxes have a scatter of 5\%. 

\section{Velocity-resolved reverberation mapping}
\begin{figure*}[ht!]
\centering
\includegraphics[angle=0,width=0.8\textwidth]{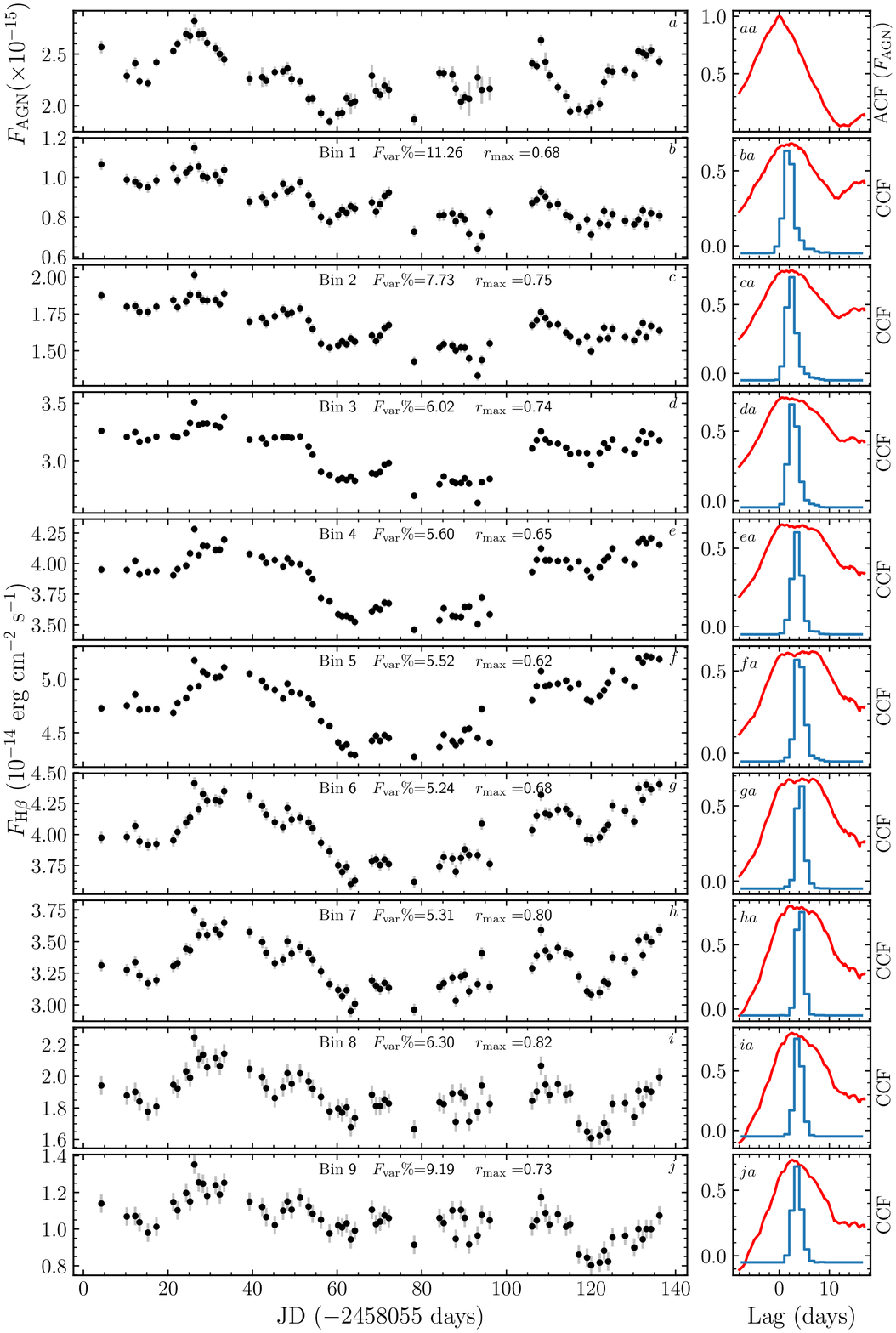}
\caption{\footnotesize
Velocity-resolved reverberation mapping. 
The left panels ({\it a-j}) show the light curves of AGN continuum at 5100~\AA~
and the broad $\rm H\beta$ emission line of each velocity bin, respectively. 
We noted the variability amplitude of the light curves in panels ({\it b-j}). 
The right panels ({\it aa-ja}) correspond to the ACF of continuum at 5100~\AA~and the
CCF between the light curve of each velocity bin ({\it b-j}) and the continuum variation ({\it a}), respectively. 
We noted the maximum correlation coefficients ($r_{\rm max}$) in panels ({\it b-j}). 
Monte Carlo simulations of the centroid (blue) are over-plotted in panels ({\it ba-ja}). 
{\tt Bin} number (Bin 1$-$9) is one-to-one correspondence with Figure~\ref{fig_vrm}. 
}
\label{fig_hbeta}
\end{figure*}

\begin{figure*}[ht!]
\centering
\includegraphics[angle=0,width=0.8\textwidth]{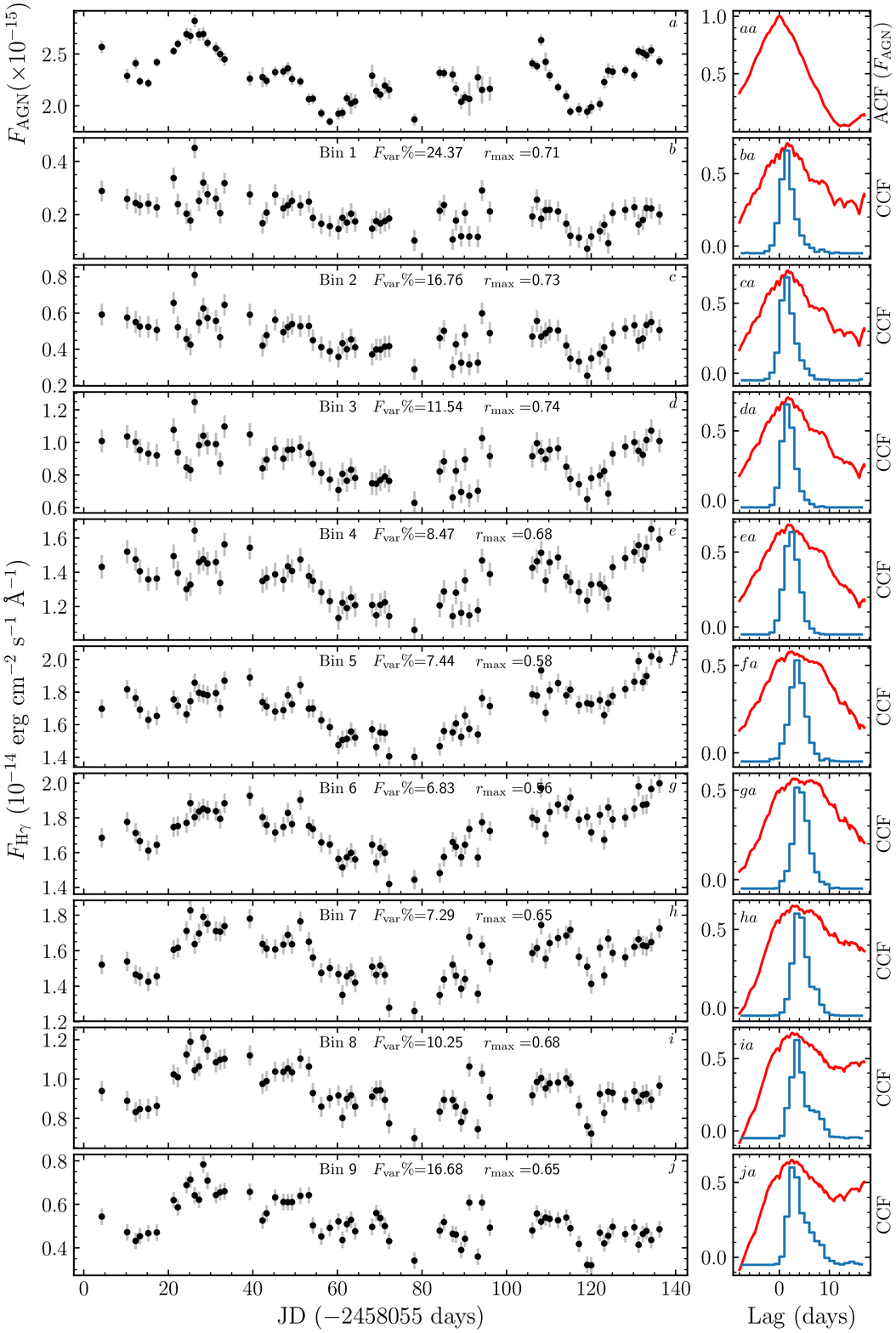}
\caption{\footnotesize
Same as Figure~\ref{fig_hbeta}, but for the broad H$\gamma$ emission line. 
}
\label{fig_hgamma}
\end{figure*}

\begin{figure*}[ht!]
\centering
\includegraphics[angle=0,width=0.8\textwidth]{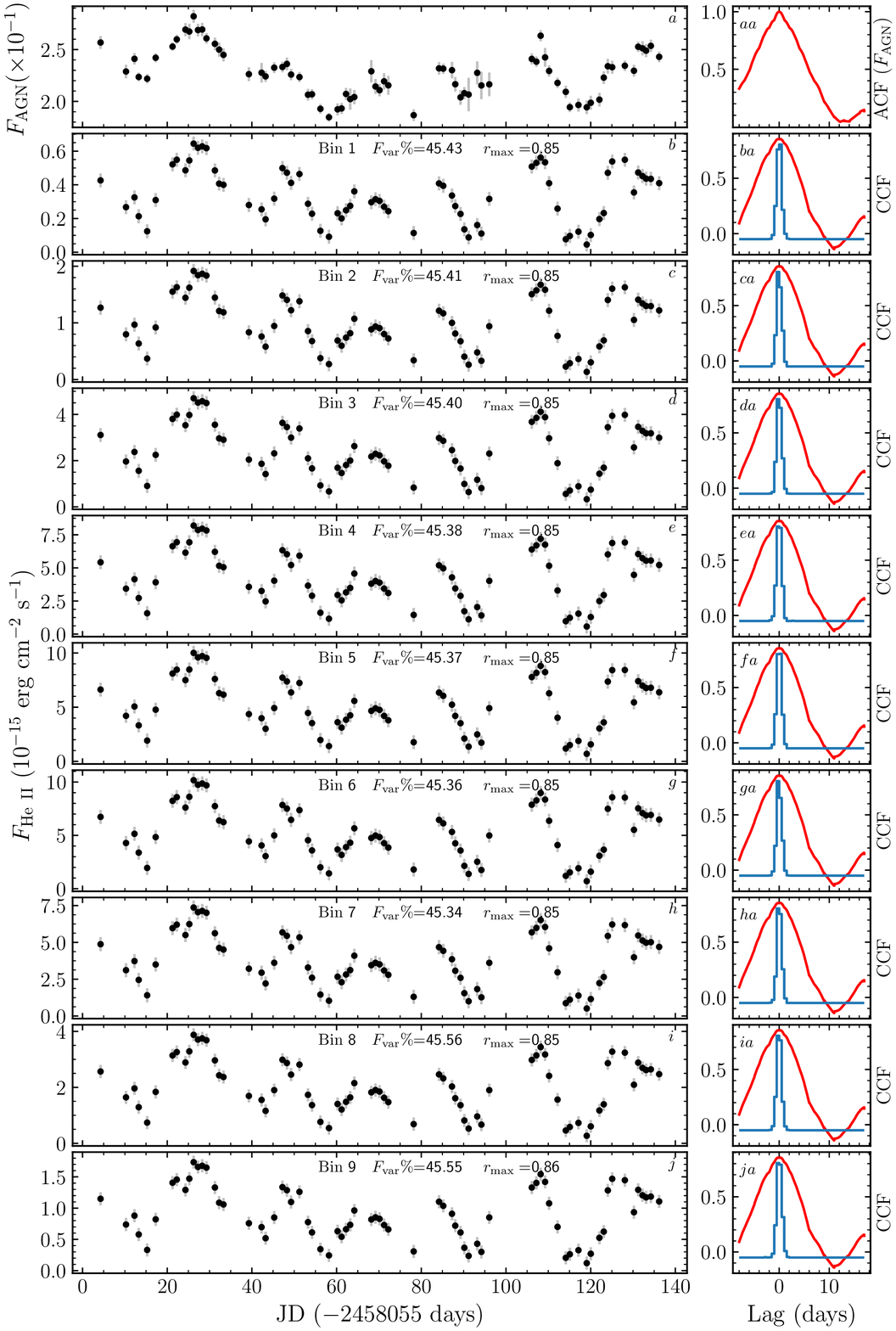}
\caption{\footnotesize
Same as Figure~\ref{fig_hbeta}, but for the broad He~{\sc ii} emission line. 
}
\label{fig_heii}
\end{figure*}

\begin{figure*}[ht!]
\centering
\includegraphics[angle=0,width=0.8\textwidth]{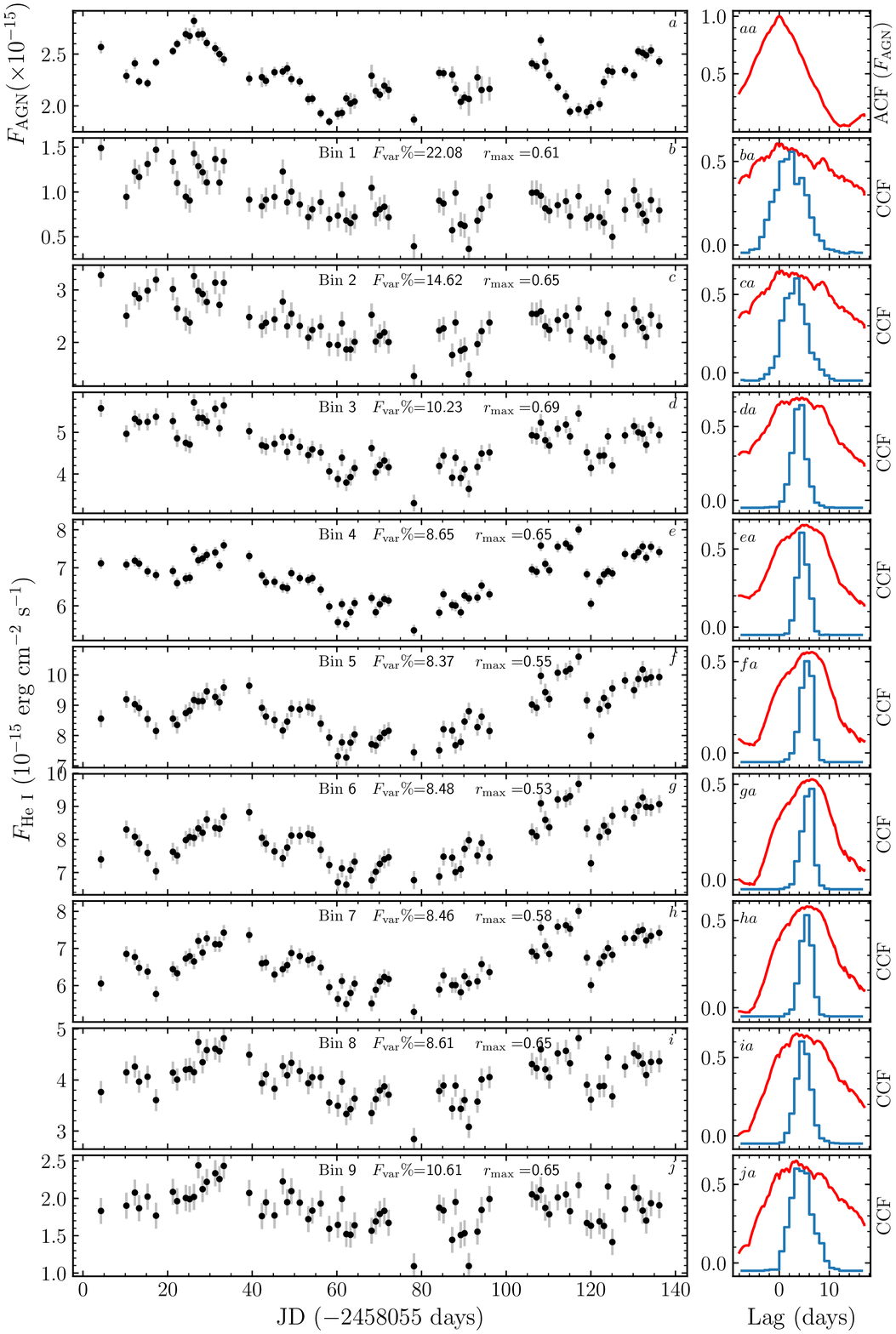}
\caption{\footnotesize
Same as Figure~\ref{fig_hbeta}, but for the broad He~{\sc i} emission line. 
}
\label{fig_hei}
\end{figure*}

In Section~\ref{sec_vm}, we presented the procedure of velocity-resolved reverberation mapping. 
In this appendix, we provide the velocity-dependent light curves and cross correlation analysis 
including H$\beta~\lambda4861$, H$\gamma~\lambda4340$, He~{\sc ii}~$\lambda4686$ and He~{\sc i}~$\lambda5876$ 
(see Figures~\ref{fig_hbeta}, \ref{fig_hgamma}, \ref{fig_heii}, \ref{fig_hei}) 

\end{document}